\documentclass[graybox]{svmult}

\usepackage{mathptmx}       
\usepackage{helvet}         
\usepackage{courier}        
\usepackage{type1cm}        
\usepackage{makeidx}         
\usepackage{graphicx}        
\usepackage{multicol}        
\usepackage[bottom]{footmisc}
\usepackage{amssymb,amsmath}
\usepackage{wasysym}
\usepackage{textcomp}
\usepackage{setspace}
\usepackage{bm}

\def\eref#1{(\ref{#1})}

\def\TL{T_{\mathrm{L}}}
\def\TR{T_{\mathrm{R}}}

\newcommand{\LL}{ {\rm L} }
\newcommand{\RR}{ {\rm R} }
\newcommand{\bra}[1]{{\langle #1 |}}
\newcommand{\ket}[1]{{| #1 \rangle}}

\newcommand{\tr}{\mathrm{tr}\, }
\newcommand{\ave}[1]{{\langle #1\rangle}}

\newcommand{\CC}{ {\rm C} }
\newcommand{\HH}{ {\rm H} }

\makeindex             

\begin{document}

\title*{From thermal rectifiers to thermoelectric devices}
\author{Giuliano Benenti, Giulio Casati, Carlos Mej\'{\i}a-Monasterio, 
and Michel Peyrard}
\institute{Giuliano Benenti \at 
Center for Nonlinear and Complex Systems,
Dipartimento di Scienze e Alta Tecnologia,
Universit\`a degli Studi dell'Insubria, via Valleggio 11, 22100 Como, Italy
and Istituto Nazionale di Fisica Nucleare, Sezione di Milano,
via Celoria 16, 20133 Milano, Italy, \email{giuliano.benenti@uninsubria.it}
\and
Giulio Casati \at
Center for Nonlinear and Complex Systems,
Dipartimento di Scienze e Alta Tecnologia,
Universit\`a degli Studi dell'Insubria, via Valleggio 11, 22100 Como, Italy
and International Institute of Physics, Federal University of Rio Grande 
do Norte, Natal, Brasil, \email{giulio.casati@uninsubria.it}
\and
Carlos Mej\'{\i}a-Monasterio \at
Laboratory of Physical Properties, 
School of Agricultural, Food and Biosystems Engineering,
Technical University of Madrid, 
Av. Complutense s/n, 28040 Madrid, Spain,
\email{carlos.mejia@upm.es}
\and
Michel Peyrard \at
Ecole Normale Sup\'erieure de Lyon,
Laboratoire de Physique CNRS UMR 5672,
46 all\'ee d'Italie, 69364 Lyon Cedex 7, France,
\email{michel.peyrard@ens-lyon.fr}}
%
%
\maketitle

\abstract{We discuss thermal rectification and thermoelectric 
energy conversion from the perspective of nonequilibrium 
statistical mechanics and dynamical systems theory. 
After preliminary considerations on the dynamical foundations of 
the phenomenological Fourier law in classical and quantum mechanics,
we illustrate ways to control the phononic heat flow and design  
thermal diodes. Finally, we consider the coupled transport of 
heat and charge and discuss several general mechanisms for 
optimizing the figure of merit of thermoelectric efficiency.}

\section{Dynamical Foundations of Fourier law}
\label{sec:fourier law}


The possibility to manipulate the heat current represents a fascinating challenge for the future,
especially in view of the need of future society of providing a  sustainable supply of energy and due to the strong concerns about the environmental impact of the combustion of fossil fuels. However  along these lines there are severe difficulties both of theoretical and experimental nature. In particular it turns out that manipulation of the heat current is much more difficult than the manipulation of the electric current.

It is therefore necessary to start from first principles in order to get a deep and systematic understanding of the properties of heat transport. Namely we would like to understand these properties starting from the microscopic dynamical equations of motion.

Along these lines a necessary step is the derivation of the Fourier heat law from dynamical equations of motion. In particular we would like to understand under what conditions Fourier law is valid.
What are the dynamical properties needed to have normal
transport in a given system? This is a nontrivial
question and for many years it has been addressed according
to different perspectives. It concerns, on one hand,
the foundations of nonequilibrium statistical mechanics
and, on the other hand, the practical issue of constructing
microscopic models which agree with the macroscopic
equations which describe transport.
For example, for a class of hyperbolic
systems (transitive Anosov) a guiding principle
was proposed (the so-called chaotic hypothesis \cite{Gallavotti})
as a prescription for extending equilibrium methods to
nonequilibrium situations. We remark that in these works
the randomness needed to obtain a consistent description
of the irreversible macroscopic phenomena comes
from the exponential instability of the microscopic chaotic
dynamics.

\begin{figure}[!h]
\centerline{\includegraphics*[width=0.7\textwidth]{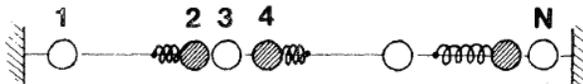}}
  \caption{The ding-a-ling model. Here the springs merely symbolize the harmonic restoring force.}
  \label{fig:dingling}
\end{figure}

First numerical evidence of the validity of Fourier heat conduction law in an exponentially unstable system was reported in \cite{Ford}
where the so-called “ding-a-ling” model was considered (Fig.~\ref{fig:dingling}). The model consists of harmonic oscillators which exchange their energy via elastic, hard-core collisions, with intermediate hard spheres.
The even-numbered particles in Fig.~\ref{fig:dingling} form a set of equally spaced lattice oscillators with each oscillator being harmonically bound to its individual lattice site and with  all oscillators vibrating at the same frequency $\omega$. The odd-numbered particles are free particles constrained only by the two adjacent even-numbered oscillators.
It can be shown that the dynamics of this model is uniquely determined by the parameter $\omega^2/E$ (where $E$ is the energy per particle)
and that the dynamics becomes exponentially unstable when this parameter is $\gg1$.
The validity of Fourier law was established in the standard way by putting the two end particles in contact with thermal reservoirs, taken as Maxwellian gases, at different temperatures. The system was then numerically integrated until the stationary state was reached and the energy exchange at the left and right reservoir became equal. This gives the average heat flux $j$. After defining the particle temperature to be twice its average kinetic energy, the value of the steady-state internal temperature gradient $\nabla T$ was computed. Then the thermal conductivity $\kappa$ was computed via the heat Fourier law $ j =- \kappa {\nabla}T$. A normal conductivity independent on the system length was 
found.

It is important to stress however that hard chaos with exponential instability is not a necessary condition to induce normal transport properties.  Moreover rigorous results are still lacking
and in spite of
several efforts, the connection between Lyapunov exponents,
correlations decay and diffusive properties is still not completely
clear. As a matter of facts it turns out that mixing property is sufficient to ensure normal heat transport \cite{Jiao}.
This might be an important step in the general
attempt to derive macroscopic statistical laws from the underlying
deterministic dynamics. Indeed, systems with zero
Lyapunov exponent have zero algorithmic complexity and, at
least in principle, are analytically solvable.

A particular case is given by total momentum conserving systems which typically exhibit anomalous conductivity.
This type of systems is largely discussed in other contributions of this volume and therefore will not be considered here ay longer. Here we would like to add only a word of caution, and to suggest that anomalous behavior in such systems is perhaps more general than so far believed.
The point is that
our present understanding of the heat conduction problem is mainly based on numerical empirical evidence while rigorous analytical results are difficult to obtain.
Numerical analysis consists of steady state, nonequilibrium simulations or of equilibrium simulations based on linear response theory and Green-Kubo formula. Typically, if both methods give reasonable evidence for Fourier law and if, moreover, they lead to the same numerical value of the coefficient of thermal conductivity $\kappa$, then this is generally considered as an almost conclusive evidence that Fourier law is indeed valid.

\begin{figure}[!h]
\centerline{\includegraphics*[width=0.7\textwidth]{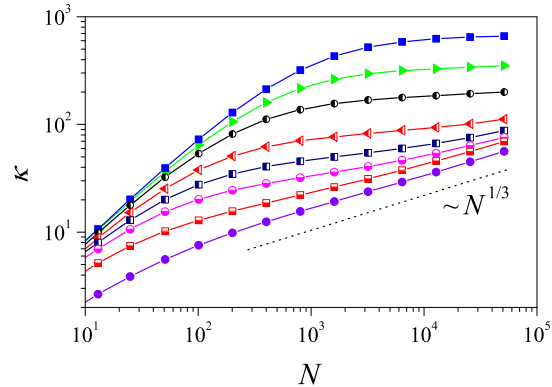}}
\caption{The heat conductivity $\kappa$ versus the
system size $N$ for the one-dimensional diatomic gas model, with alternative 
mass $M$ and $m$. From top to bottom, the
mass ratio $M/m$ is respectively $1.07$, $1.10$, $1.14$, $1.22$, $1.30$,
$1.40$, the golden mean ($\approx1.618$), and $3$.}
\label{fig:fourierlike}
\end{figure}

This conclusion, however, might be not correct as shown in \cite{Chen2014}, where the heat conductivity of the one-dimensional diatomic hard-point gas model was studied. 
As shown in Fig.~\ref{fig:fourierlike}, the Fourier-like behavior,
seen in both equilibrium and nonequilibrium simulations, turns out to be  a finite-size effect and Fourier law appear to hold up to some size $N$ after which anomalous behavior sets in. This behavior requires a better understanding. Indeed,
while it is natural to expect an initial ballistic behavior for larger and larger $N$ as one approaches the integrable limit, it is absolutely not clear why the value of $\kappa$ appears to saturate to a constant value before entering the anomalous regime $\kappa \sim N^\alpha$ (with $\alpha\approx 1/3$) 
at an even larger system size $N$.

To summarize, while
establishing a complete connection between ergodic properties
and macroscopic transport features is still beyond the
reach of present understanding, we may conclude that apart some particular notable exceptions, dynamical mixing property induces  deterministic diffusion
and hence Fourier law.

\section{Fourier law in quantum mechanics}
\label{sec:QFL}

The next step is to discuss whether or not Fourier heat law can be derived
from quantum dynamics without a priori statistical assumptions. This calls
directly in question the issue of `Quantum chaos''.
The  first  attempts to  provide  a  microscopic  description of  heat
transport in quantum  systems dates back to the  beginning of the 20th
century  with   the  work  of  Debye  in   1912  \cite{Debye1912}  and
subsequently  of   Peierls  in  1929   \cite{Peierls1929}.   Based  on
modifications  of the  kinetic Boltzmann  equation these  theories are
classical  in essence  by considering  classical-like quasi-particles,
and  fail  to  describe  systems  out of  equilibrium  with  dimensions
comparable to the electron and phonon mean free paths.

The recent achievements in the miniaturisation of devices have boosted
the  interest in  understanding  the conditions  under  which heat  is
transported  diffusively in  quantum systems.   In spite  of  the many
efforts a  rigorous derivation  of a quantum  Fourier law  for general
Hamiltonians remains  an unsolved problem.  A main  difficulty to the
study  of  heat  conduction  in  quantum  mechanics  is  the  lack  of
appropriate definitions  of local  quantities such as  the temperature
and  the  heat current  \cite{Dubi2009}, and  calls  in question  the
problem of thermalization,  namely the relaxation to a  state in local
equilibrium,  in isolated  \cite{Aberg1990} and  open  quantum systems
\cite{Breuer2007}.  It   has  been  found  that   the  conditions  for
thermalization are  essentially related to  the systems' integrability
and  localization  properties  (e.g.  due  to  disorder).  Non-ergodic
systems,   undergo   relaxation   to   a   generalized   Gibbs   state
\cite{Barthel2008},  so  that  the  application  of  standard  statistical
mechanics methods is possible.

Quantum systems in contact with  external heat baths can be treated by
using      the      Lindblad-Gorini-Kossakowski-Sudarshan     equation
\cite{Lindblad1976,Gorini1976}  in a  convenient setup  in  which only
boundary degrees of freedom  are coupled with the environment.  Within
the  Markovian approximation,  the system's  many-body  density matrix
evolves according to
\begin{equation}
\frac{d}{dt}\rho(t) = \hat{\cal L}\rho(t),
\label{eq:lind}
\end{equation}
where the Liouvillian superoperator is defined as
\begin{equation}
\hat{\cal L}\rho := -\frac{i}{\hbar}\,[\mathcal{H},\rho] + \sum_{\mu} \left(L_\mu  \rho L^\dagger_\mu - \frac{1}{2}\,\{L_\mu^\dagger L_\mu,\rho\}\right).
\end{equation}
We assume here that the  Hamiltonian $\mathcal{H}$ can be written as a
sum  of  locally  interacting  terms, $\mathcal{H}=\sum_{n}  H_n$  and
$L_\mu$ are the Lindblad  (or so-called quantum jump) operators, which
are assumed  to act  only at  the boundary sites  of the  system. This
setup provides a fully  coherent bulk dynamics and incoherent boundary
conditions, which is  particularly suited for studying nonequilibrium
heat transport in a setup similar to the classical case \cite{Wichterich2007}.

The Quantum  Master Equation (QME) approach  can be used  to study not
only   heat  transport,  but   also nonequilibrium  processes   in  general
(particle transport, spin transport,  {\it etc.}).  Depending  on the process  in question
the  Lindblad  operators  $L_\mu$  target specific  canonical  states,
creating a local equilibrium state  near the boundaries of the system.
The conductivities  are then obtained by  measuring expectation values
of  the current  observables  in  the steady  states  of the  Lindblad
equation,   in  the   thermodynamic  limit   $N\to\infty$   (see  {\it
  e.g.}, \cite{Prosen2009}).  This  approach has been  extensively used
in  recent  years to  study  heat  transport   in
one-dimensional models  of quantum spin  chains coupled at  their ends
with                 Lindblad                heat                baths
\cite{Dubi2009,Wu2008,Michel2008,M-M2007,Manzano2012,Sun2010,Ajisaka2012},
as     well     as     in     chains    of     quantum     oscillators
\cite{Zunkovic2012,Gaul2007}  (for  a recent  review  see {\it  e.g.},
\cite{Michel2006,TEreview}).

The  Lindblad equation (\ref{eq:lind})  allows  efficient numerical
simulation  of the  steady state  of locally  interacting  systems, in
terms   of   the   time-dependent-density-matrix-renormalization-group
method  (tDMRG)  \cite{Daley2004,White2004,Uli2011}  in the  Liouville
space of linear operators  acting on wave functions \cite{Prosen2009}.
In  cases where  the tDMRG  method cannot  be applied,  like  when the
interaction is long-range ({\it e.g.}, Coulomb), the QME can be
solved using  the method of  quantum trajectories, (see,  for example,
\cite{M-M2007}).  In the latter case the idea is to represent the density operator
as  an  expectation  of  $\ket{\Psi}\bra{\Psi}$  where  the  many-body
wave function  $\Psi$  is  a  solution of  a  stochastic  Schr\"odinger
equation $d\Psi(t) = -(i/\hbar) H\Psi dt + d\xi$, with $d\xi$ being an
appropriate stochastic process simulating the action of the baths.  In
addition, this method has the advantage that non-Markovian effects can
be treated easily and intuitively.
In more general settings, the QME can always be solved exactly through
numerical integration where the quantum canonical heat baths are often
modeled in  terms of the Redfield  equation \cite{Redfield1957}.  Such
approach has found a broad  applicability in many-body systems and has
been      used     to      investigate      heat     transport~\cite{Saito2000,Saito2003,Segal2005,Wu2008}.

One alternative of using the  QME approach is the Keldysh formalism of
nonequilibrium Green's functions,  where one essentially discusses the
scattering  of elementary  quasi-particle excitations  between  two or
more infinite non-interacting Hamiltonian reservoirs. 
The  Keldysh formalism considers  an initial product
state  density matrix describing  the finite  system and  two infinite
baths    in   thermal   equilibrium    at   {\it    e.g.},   different
temperatures. The system and the  reservoirs are then coupled and the
density matrix is  evolved according to the full  Hamiltonian.  In the
steady state, currents and local densities can be obtained in terms of
the so-called Keldysh Green’s functions.  This approach has been used,
among  other  things, to  study  heat  transport  in driven  nanoscale
engines  \cite{Arrachea2007,Arrachea2012}  and  spin  heterostructures
\cite{Arrachea2009}.
Another  commonly used  approach to  study heat  transport  in quantum
systems is  based on the  Green-Kubo formula, originally  developed to
study electric transport \cite{Kubo1957}.

Within linear  response theory, the  current is taken as  the system's
response to  an external perturbative potential which  can be included
within the  Hamiltonian of the  system. First order  perturbation theory
yields the Green-Kubo formula relating the nonequilibrium conductivity
with  the  equilibrium current-current  correlation.  This formula  is
naturally  extended to study  heat transport,  where the  heat current
appears as the response to an external temperature gradient. This {\it
  ad-hoc} generalization remains  conceptually troublesome since there
is  no  potential term  in  the  system's  Hamiltonian representing  a
temperature                     gradient                     situation
\cite{Kubo1957,Mahan1990}. In spite of this, the Green-Kubo approach
has become a widely employed \cite{Zotos1997,Zotos2004,Garst2001,H-M2005}.

In spite  of all efforts, a  microscopic derivation of  Fourier law in
quantum  mechanics is  still lacking,  and only  partial understanding
concerning the  conditions under  which this is  expected to  hold has
been  gained.   Particularly,  in  analogy  with  the  studies  at  the
classical level the  relation between the validity of  Fourier law and
the  onset of  quantum chaos  has  been investigated  in recent  years
\cite{Saito2000,Michel2003,M-M2005,Steinigeweg2006,Prosen2009,Prosen2010}.

As it  has been shown in  the previous section  for classical systems,
diffusive heat transport is directly  related to the chaoticity of the
dynamics. While  such relation is not  strict, classical deterministic
chaos is  yet expected to yield diffusive  behavior.
It is  nowadays well established  that quantum
systems  for  which their  classical  analogues  are chaotic,  exhibit
characteristic signatures  in the spectra and  the eigenfunctions that
are  different from  those observed  in systems  that  are classically
integrable \cite{Bohigas1984,Casati1980}.  The global manifestation of
the  onset of  chaos in  quantum systems  consists of  a  very complex
structure of  the quantum states  as well as in  spectral fluctuations
that   are   statistically   described   by   Random   Matrix   Theory
\cite{Guhr1998}. In the following we discuss
the relation among the validity of quantum Fourier law and the onset of quantum chaos.

\subsection{Fourier law and the onset of quantum chaos}
\label{sec:QC}

\begin{figure}[!h]
  \centerline{\includegraphics*[width=0.65\textwidth]{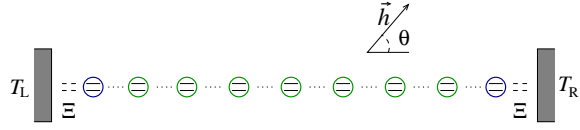}}
  \caption{The finite quantum spin chain model coupled to external
    heat baths at different temperatures. The dotted lines represent
    the nearest neighbour interaction. The double dashed lines
    represent the coupling $\Xi$ with the baths. The angle $\theta$ of
    the magnetic field is measured with respect to the direction $x$ of
    the chain.}
  \label{fig:qfl-model}
\end{figure}

The relation between  the validity of the quantum  Fourier law and the
onset  of quantum chaos  in a  genuinely nonequilibrium  situation was
studied  in \cite{M-M2005}.  There a quantum Ising  chain of  $N$ spins
$1/2$ subjected to a  uniform magnetic field
$\vec{h} = (h_x,0,h_z)$ and coupled at its extremes with quantum heat
baths, was considered.
The Hamiltonian of this system is
\begin{equation} \label{eq:H}
\mathcal{H} = \sum_{n=0}^{N-2}H_n +
\frac{h}{2}(\sigma_\mathrm{L} + \sigma_\mathrm{R}) \ ,
\end{equation}
where $H_n$ are local energy density
operators appropriately defined as
\begin{equation} \label{eq:H_local}
H_{n}     =   -Q\sigma^z_n\sigma^z_{n+1}    +
\frac{\vec{h}}{2} \cdot \left(\vec{\sigma}_n  + \vec{\sigma}_{n+1}\right) \ ,
\end{equation}
and  $\sigma_\mathrm{L} = \vec{h}\cdot\vec{\sigma}_0/h$,  $\sigma_\mathrm{R} =
\vec{h}\cdot\vec{\sigma}_{N-1}/h$ are  the spin operators  along the direction
of  the magnetic  field of  $s_0$ and  $s_{N-1}$ respectively.   The operators
$\vec{\sigma}_n =  (\sigma^x_n,\sigma^y_n,\sigma^z_n)$ are the  Pauli matrices
for the $n$-th  spin, $n=0,1,\ldots N-1$. A schematic representation
of this model is shown in Fig.~\ref{fig:qfl-model}.

In this  model, the angle $\theta = \arctan (h_z/h_x)$ 
of  the magnetic field makes  with the chain
affects  the dynamics of the system. If
$\theta=0$,  the  Hamiltonian (\ref{eq:H})  corresponds  to the  Ising
chain  in  a  transversal  magnetic  field,  which  is  integrable  as
(\ref{eq:H})  can be  mapped into  a  model of  free fermions  through
standard Wigner-Jordan  transformations. For $\theta >  0$, the system
is no  longer integrable and for $\theta\approx  \pi/4$, quantum chaos
sets   in.  The   system  becomes   again  (nearly)   integrable  when
$\theta\approx \pi/2$.  Therefore, by tuning $\theta$  one can explore
different regimes of quantum dynamics and study the relation between
the integrability of the system and the validity of Fourier's law.

The  integrability of  a quantum  system can  be characterised  by the
Nearest  Neighbour Level  Spacing  (NNLS) distribution  $P(s)$,
which is  the
probability density to find two adjacent levels at a distance $s$. For
an integrable  system the distribution $P(s)$ has  typically a Poisson
distribution:
\begin{equation}\label{eq:pds_P}
P_{\rm P}(s)=\exp\left(-s\right) \ .
\end{equation}
In  contrast, in the  quantum chaos  regime,  Hamiltonians obeying
time-reversal  invariance exhibit a NNLS distribution that
corresponds  to the  Gaussian Orthogonal  Ensemble of  random matrices
(GOE).  This distribution is  well-approximated by the Wigner surmise,
which reads
\begin{equation} \label{eq:pds_WD}
P_{\rm WD}(s) = \frac{\pi s}{2}\exp\left(-\frac{\pi s^2}{4}\right) \ ,
\end{equation}
exhibiting ``level repulsion''.

\begin{figure}[!h]
  \centerline{\includegraphics*[width=0.70\textwidth]{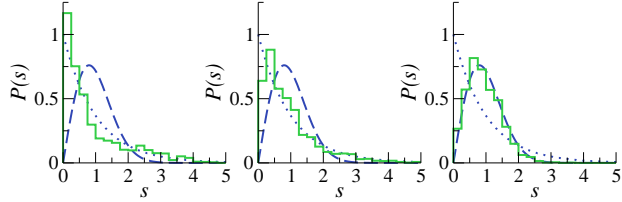}}
  \caption{NNLS distribution $P(s)$ for the integrable (left panel),
    intermediate (middle panel) and chaotic (right panel) spin
    chains. The histogram was numerically obtained for a chain of $N=12$
    spins by diagonalizing Hamiltonian (\ref{eq:H})
and averaging over the spectra of even and odd
    parity. The dotted  curve corresponds to $P_{\mathrm{P}}$ and the
    dashed curve to $P_{\mathrm{WD}}$.}
  \label{fig:qfl-Ps2}
\end{figure}

Figure~\ref{fig:qfl-Ps2} shows the NNLS distribution $P(s)$ for three different
directions  of the  magnetic field:  ({\it i})  {\em  integrable case}
$\vec{h}=(3.375,0,0)$, at  which $P(s)$  is well described  by $P_{\rm
  P}(s)$, ({\it ii})  {\em intermediate case} $\vec{h}=(7.875,0,2)$ at
which  the distribution  $P(s)$ shows  a combination  of  (weak) level
repulsion and  exponential decay, and  ({\it iii}) {\em  chaotic case}
$\vec{h}=(3.375,0,2)$  at which  the distribution  $P(s)$  agrees with
$P_{\rm WD}(s)$ and thus corresponds to the regime of quantum chaos.

In  Ref.~\cite{M-M2005} a numerical  method to  solve the  dynamics of
open  quantum spin  chains was  introduced.  This  method  consists in
periodically and  stochastically collapsing the state of  the spins at
the boundaries of  the chain to a state that  is consistent with local
equilibrium states at different temperatures. These stochastic
quantum  heat baths  are analogous  to  the stochastic  baths used  in
classical  simulations and  even when  this  method does  not yield  a
stochastic unravelling  of QME, it is numerically  simple to implement
and analyse (for more details see Refs.~\cite{M-M2005,M-M2007}).

\begin{figure}[!h]
  \centerline{\includegraphics*[width=0.65\textwidth]{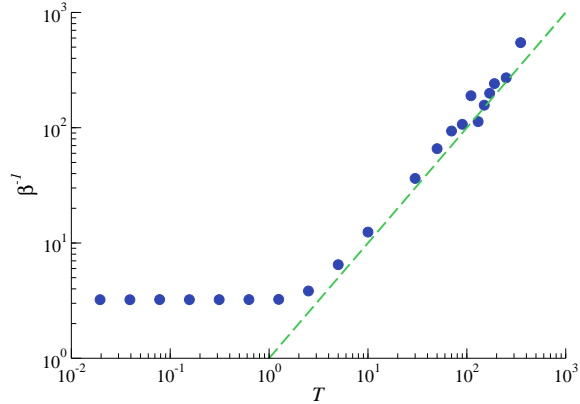}}
  \caption{Local temperature in the center of the chain $\beta^{-1}$
    as a function of the temperature of the baths $T$, obtained from
    equilibrium simulations in a chain of $7$ spins, as the best fit
    to exponential of the local density matrix $\rho_n(E_n)$ in the
    central symmetry band. The dashed line stands for the identity.}
  \label{fig:qfl-beta}
\end{figure}

 Using  this method,  local  thermal equilibrium  was  first checked  by
computing  time  averages of the density matrix of the system:
\begin{equation} \label{eq:rho}
\overline{\rho} = \lim_{t\rightarrow\infty} \int_0^t
\ket{\psi(s)}\bra{\psi(s)}{\mathrm d}s \ ,
\end{equation}
where $\psi(s)$  is the state of  the system at time  $s$. Setting
both heat baths to the same temperature $T_\LL  = T_\RR = T$, it was found that
$\overline{\rho}$ is diagonal within numerical accuracy and
consistent with
\begin{equation} \label{eq:rhoeq}
\bra{\phi_n}\rho\ket{\phi_m} = \frac{e^{-\beta E_n}}{\mathcal{Z}}
\delta_{m,n} \ ,
\end{equation}
inside each symmetry band.  Here $\ket{\phi_n}$ are the eigenfunctions
in  the  energy  basis, $\mathcal{H}\ket{\phi_n}=E_n\ket{\phi_n}$,  and
$\mathcal{Z}  =  \sum_n e^{-\beta  E_n}$  is  the canonical  partition
function. From a best fit to exponential of Eq.~(\ref{eq:rhoeq}) a value
of the local  temperature in the bulk of the  system can be extracted.
The results are shown in Fig.~\ref{fig:qfl-beta} as a function of the
temperature of the heat baths. It can be seen that for large enough
temperatures of the heat baths ($T\gtrsim 5$) the system thermalizes to exactly the
same temperature \cite{M-M2007}.

Out  of equilibrium  expectation values  in the  nonequilibrium steady
state  were obtained  as  follows: for  each  realization, the  initial
wave function $\ket{\psi(0)}$  of the system  is chosen at  random. The
system is then evolved for some relaxation time $\tau_{\rm rel}$ after
which  it  is assumed  to  fluctuate  around  a unique  steady  state.
Measurements are  then performed as  time averages of  the expectation
value of  the observables, that  are further averaged  over different
random realizations.

\begin{figure}[!h]
  \centerline{\includegraphics*[width=0.65\textwidth]{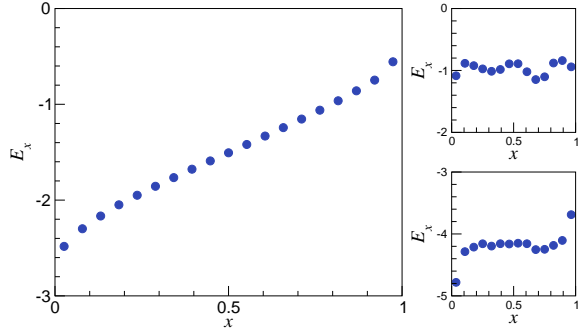}}
  \caption{Energy profile obtained from a time average of the
    expectation value of the energy density operator $E_x = \langle
    H_x\rangle$, with $x = n/N$ for a chain of $20$ spins. The
    temperatures of the baths are $\TL=5$ and $\TR=50$. The
    different panels are for the chaotic (left), intermediate (bottom
    right) and integrable (upper right) chains.}
  \label{fig:qfl-En}
\end{figure}

Figure~\ref{fig:qfl-En} shows the energy profile obtained from the time average of
the local energy density operator $E_x = \langle H_{x} \rangle$
(with $x=n/N$), for
the above three different spin chains. Interestingly, for the chaotic chain,
a linear energy profile in the bulk of the chain was found. This
indicates that the chaotic chain is able to sustain a heat current
which depends on the nonequilibrium imposed by the external heat
baths. In contrast, the integrable chain shows a flat constant energy
profile. The intermediate chain which is neither chaotic nor
integrable is not able to sustain a diffusive heat current and shows
and energy profile which is flat except near the boundaries.

To directly check the validity of Fourier's law it is possible to define
 local  heat current  operators  using  the continuity equation  for
 the  local energy  density operators  $\partial_t{H_n}  = i[{\mathcal
   H},H_n] = - (j_{n+1} - j_n)$, requiring that $j_n = [H_n,H_{n-1}]$.
From Eqs.~(\ref{eq:H}) and (\ref{eq:H_local}) the local current operators
 are explicitly given by
\begin{equation}
j_{n} = h_xQ\left(\sigma_{n-1}^z-\sigma_{n+1}^z\right)\sigma^y_{n},
\quad
1\le n\le  N-2.
\end{equation}

\begin{figure}[!h]
  \centerline{\includegraphics*[width=0.65\textwidth]{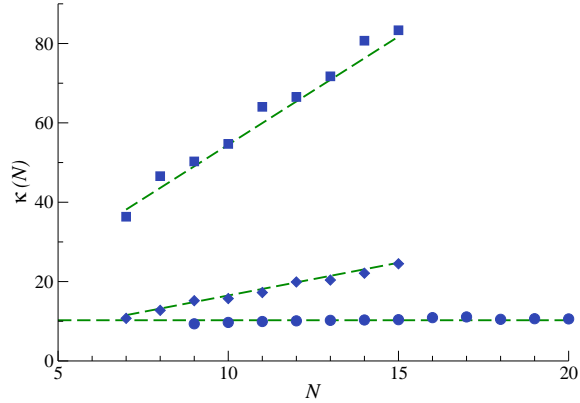}}
  \caption{Heat conductivity $\kappa=-j/\nabla T$ as a function of the
    size of the spin chain $N$, for the chaotic (circles),
    intermediate (diamonds), and integrable (squares) chains. The
    temperatures of the baths are $\TL=5$ and $\TR=50$. The
    dashed curves correspond to the best linear fit for each of
    the data sets.}
  \label{fig:qfl-kappa}
\end{figure}

Figure~\ref{fig:qfl-kappa} shows the heat  conductivity as a function of
the  system size  $N$,  calculated as  $\kappa=-j/\nabla  T$. The  mean
current $j$ was calculated as  an average of $\ave{J_n}$ over time and
over the $N-8$ central spins.  For the particular choice of the energy
density operator (\ref{eq:H_local}), its averaged expectation value is
related   to  the   local  temperature   as   $\ave{H_n}\propto  -1/T$
\cite{M-M2005}.  The  temperature  difference  was  thus  obtained  as
$\Delta T  = -1/\ave{H_{N-5}} +  1/\ave{H_3}$. For large $N$  the heat
conductivity of the chaotic chain was found to converge to a constant
value, thus confirming the validity of the Fourier's law. On the
contrary, for the integrable and an intermediate chains, $\kappa$
diverges linearly with $N$, which is a signature of ballistic transport.

These results represent a solid suggestion that,
in analogy to what is observed in classical systems, in the quantum
realm  Fourier law  holds once  quantum  chaos has  set in.  Arguably,
quantum  chaos  yields  diffusive   heat  transport  as  it  leads  to
exponential decay  of the ``dynamic'' correlations,  in particular the
energy current-current  correlation that defines  the heat conductivity
through a Green-Kubo formula.

The crucial relation between diffusive transport and quantum chaos was
later  investigated in \cite{Steinigeweg2006}  for models  of isolated
quantum chains made of  interacting subunits, each containing a finite
number of
energy    levels.    These    models    representing   single-particle
multi-channel quantum wires, exhibit  a transition to quantum chaos as
the strength  of the interaction  between the subunits  increases.  By
solving the corresponding Schr\"odinger equation, it was found that
the evolution of  the local energy density operators  is in agreement
with the corresponding diffusion equation
only  when the  system  level statistics  is
chaotic. There, a Heisenberg spin  chain in an external magnetic field
was also studied, yielding the same result.

Later, in  Ref.~\cite{Prosen2009} heat  and spin transport  in several
open  quantum spin  chains was  considered and  numerically  solved by
means of the tDMRG method. The same model considered in \cite{M-M2005}
was studied for much larger system sizes and the relation between
quantum Fourier law and quantum chaos put forward there, was
reconfirmed with high accuracy.

\section{Controlling the heat flow: thermal rectifiers}
\label{sec:rectifiers}

Contrary to the case of electronic transport, where the concept of diode
is well known, when one thinks of heat flow and Fourier law, the idea of
directed transport does not come to mind at all. It is even counter
intuitive. However, as shown in Sec.~\ref{subsec:diode}, the
concept of a thermal diode is perfectly compatible with the usual
Fourier law, provided one builds a device with materials having a
temperature-dependent thermal conductivity. With a simple
one-dimensional model system, Sec.~\ref{subsec:oned} shows how
such materials could be obtained. As shown in
Sec.~\ref{subsec:deuxd} the same results can be extended in higher
dimensions. The actual realization of thermal rectifiers is briefly
discussed in Sec.~\ref{subsec:actualrectifs}

\subsection{The Fourier law and the design of a thermal rectifier}
\label{subsec:diode}

Thermal rectification is everyday's experience: due to thermal convection
a fluid heated from below efficiently transfers heat upwards, while the
same fluid, heated from the top surface shows a much weaker transfer
rate downwards. In this case this is because the heat flow is
due to a transfer of matter. The idea that one could build a solid-state
device that lets heat flow more easily in one way than in the other is
less intuitive, and may even appear in contradiction with thermodynamics
at first examination. However this is not so, and the design of a
thermal rectifier is perfectly compatible with the Fourier law
\cite{PeyrardEPL2006}.

Let us consider the heat flow along the $x$-direction, in a material in
contact with two different heat baths at temperature $T_1$ for $x=0$ and
$T_2$ for $x = L$. A rectification can only be expected if the device
has some spatial dependence which allows us to distinguish its two ends,
i.e. if the local thermal conductivity depends on $x$. This can either
come from an inhomogeneity of the material or from its
geometry. Moreover the thermal conductivity $\kappa(x,T)$
may also depend on temperature so that the
Fourier law relates the heat flux  $j_f$ to the local temperature $T(x)$
by
\begin{equation}
  \label{eq:tdistrib}
T(x) = T_1 + \int_0^x \dfrac{j_f}{\kappa[\xi,T(\xi)]} \; d \xi \; .
\end{equation}
Solving this equation with the boundary condition $T(x=L) = T_2$
determines the value of $j_f$. If the boundary conditions are reversed,
imposing temperature $T_2$
for $x=0$ and temperature $T_1$ for $x=L$, solving the same
equation leads to
another temperature distribution and another distribution of the
local thermal conductivity $\kappa(x,T)$. Therefore the reverse flux
$j_r$ is not equal to the forward flux $j_f$.
The rectifying coefficient can be defined as
\begin{equation}
  \label{eq:defR}
R = \left| \dfrac{j_r}{j_f}\right| \; .
\end{equation}
In general, for arbitrary $\kappa(x,T)$, there is no condition that
imposes that $R$ should be unity.

\begin{figure}[!h]
\centerline{\includegraphics*[width=0.45\textwidth]{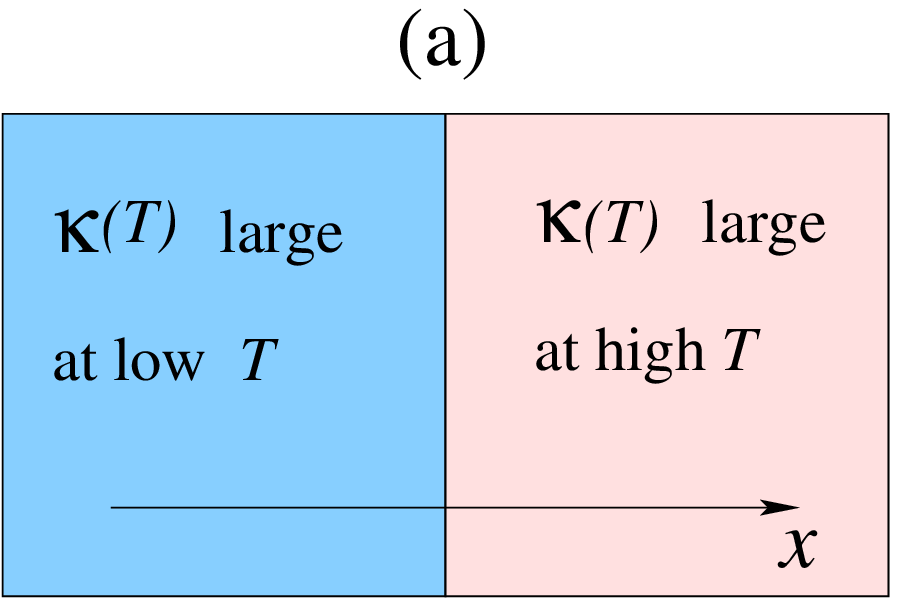}
\hspace{0.2cm}\includegraphics*[width=0.45\textwidth]{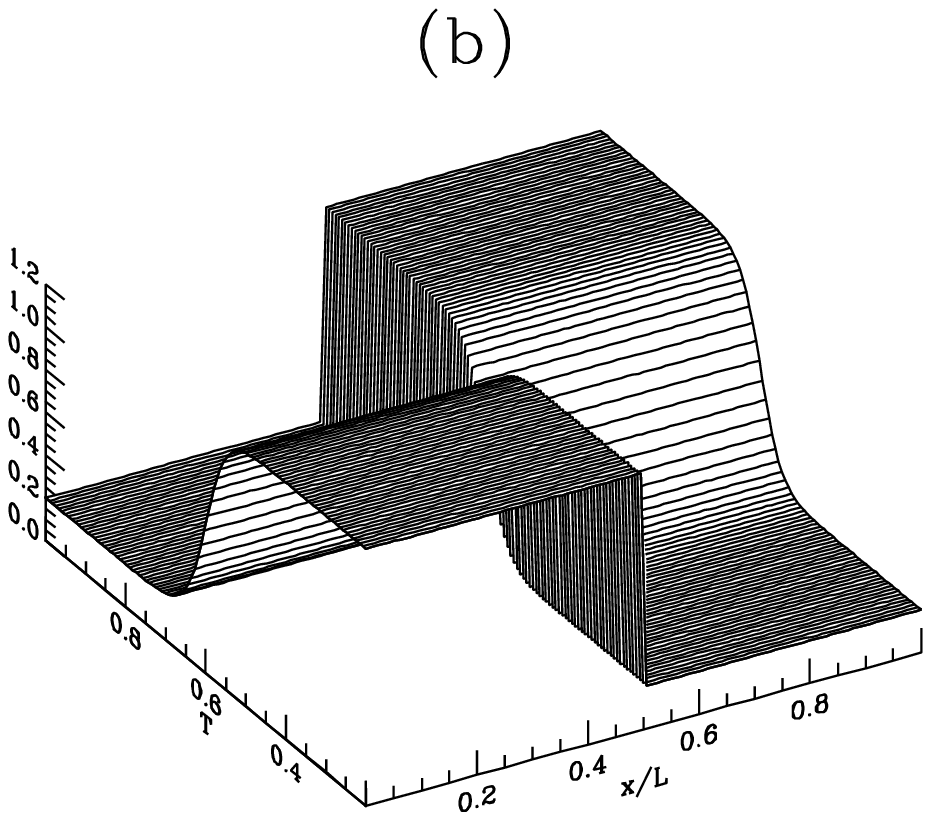}}
\vspace{0.2cm}
\centerline{\includegraphics*[width=0.45\textwidth]{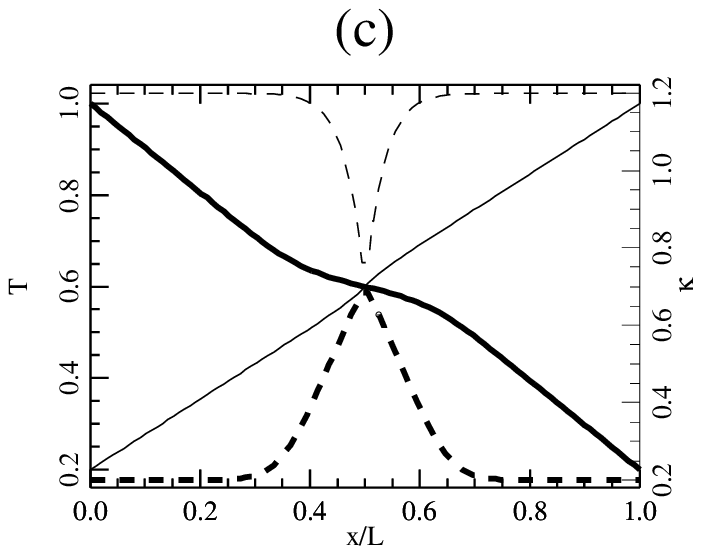}}
\caption{Thermal rectifier made by the juxtaposition of two different
  homogeneous materials which have a thermal conductivity that
  highly depends on temperature.
The boundary temperatures are $T_1 = 1.0$ and $T_2 = 0.2$ in arbitrary
  scale.
(a) Schematic view of the device.
(b) Variation of $\kappa(x,T)$. (c) The temperature distributions
  (solution of Eq.~(\ref{eq:tdistrib}))
  (full lines) and the variation versus space of the local
  conductivity $\kappa[x,T(x)]$ (dashed lines) are shown
   for the forward boundary condition
  ($T(x=0) = T_1$, $T(x=L) = T_2$) (thick lines) and reverse boundary
  condition (thin lines). For this choice of $\kappa(x,T)$, the
  rectifying coefficient is $R = |j_r/j_f|=4.75$. 
}
\label{fig:tdist2}
\end{figure}

Figure~\ref{fig:tdist2} shows a simple example where the spatial
dependence is obtained by juxtaposing two different homogenous
materials, each one
having a thermal conductivity that strongly depends on temperature. In
this case $\kappa(x,T)$ is a sigmoidal function in both cases, but on one
side $\kappa$ is large at low temperature while, on the other side it
is large at high temperature. An even simpler device can be
obtained by combining one material with a temperature dependent thermal
conductivity with another one which has a constant thermal
conductivity \cite{PeyrardEPL2006}. Such a device has a lower rectifying
coefficient but nevertheless behaves as a thermal diode.

\subsection{A one-dimensional model for a thermal rectifier}
\label{subsec:oned}

As shown in Sec.~\ref{subsec:diode}, in order to obtain a thermal
rectifier, we need two basic ingredients, a temperature dependent
thermal conductivity and the breaking of the inversion symmetry of the
device in the direction of the flow. In this section we show how this
can be obtained in a simple model system.

\begin{figure}[!h]
\centerline{\includegraphics*[width=0.6\textwidth]{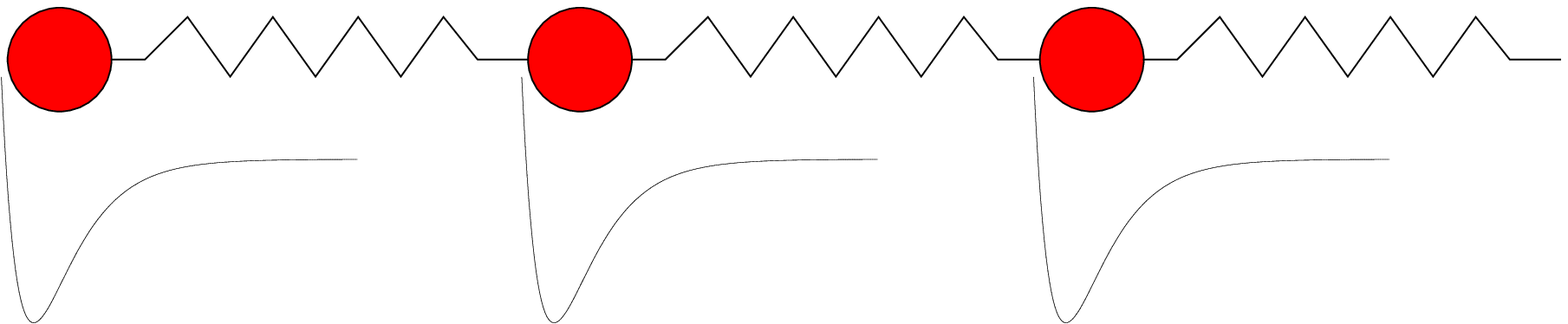}}
\vspace{0.2cm}
\centerline{\includegraphics*[width=0.6\textwidth]{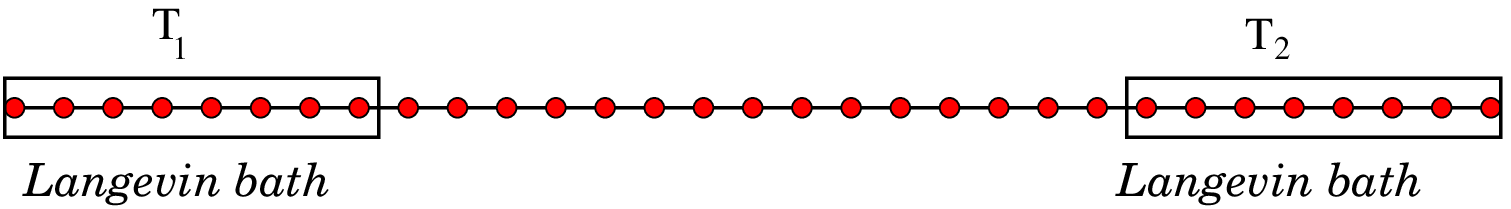}}
\caption{Schematic picture of the one-dimensional model used as the
  basis for a simple thermal rectifier. Upper part: the harmonically
  coupled particles are subjected to an on-site potential, here a Morse
  potential. Lower part: the model used in numerical simulations to
  measure the heat flow. The two end-segments (boxes) are in contact with
  a numerical Langevin thermostat, while the central part of the lattice
is evolving according to the equations of motions that derive from the
Hamiltonian (\ref{PBmodel}).}
\label{figonedmodel}
\end{figure}

In a solid the heat transfer by conduction is a transfer of energy without a
transport of matter. Heat can be carried either by the propagation of atomic
vibrations, i.e.\ phonons, or by the diffusion of the random
fluctuations of mobile particles,
which are generally charged so that electrical and heat conductivity are
closely related as stated by the Wiedemann-Franz law for metals
\cite{ashcroft1976}. Here we consider the case of electrical insulators in
which heat is only carried by lattice vibrations. The simplest model of
a thermal diode can be designed with a one dimensional lattice of
interacting particles having a single degree of freedom. However, in the
search of simplicity, one should make sure that the model does not lead
to unphysical properties. In particular we want to select a model system
that obeys the macroscopic Fourier law, with a well defined thermal
conductivity $\kappa$, which may not be the case for a one-dimensional
lattice \cite{Lepri2003}. However if the translational invariance
is broken by a substrate potential, so that momentum is not a constant
of the motion, a simple one-dimensional lattice of harmonically coupled
particles subjected to an external potential, known as a Klein-Gordon
model, can show a well defined thermal conductivity while allowing an
easy analysis of the properties of the system. Such a lattice can form
the basis for a thermal rectifier \cite{TerraneoPRL2002}.

As an example, let us consider the model schematized in
Fig.~\ref{figonedmodel}, {\it i.e.} a chain of $N$ particles with harmonic
coupling constant $K$ and a Morse on-site potential
$V_n = D_n \big[ \exp (-\alpha_n  y_n) -1 \big]^2  $. The variable $y_n$
designates the displacement of the particles with respect to their
equilibrium positions, $p_n$ their momentum, and $H_n$ is the
local energy density. This model was introduced as a  simple
one-dimensional model of DNA \cite{PBD}. In this case the on-site
potential describes the interaction between the two strands of DNA.

In the present context this model can simply be viewed as a simple example
to study heat transfer in a one-dimensional lattice, with Hamiltonian
\begin{equation}
\mathcal{H} = \sum_{n=1}^{N} H_n = \sum_{n=1}^{N} \left[\frac{p_n^2}{2 m}
+ \frac{1}{2} K
(y_n-y_{n-1})^2 
+ D_n(e^{-\alpha_n  y_n}-1)^2\right].
\label{PBmodel}
\end{equation}
In such a system we can define a local temperature by $T_n = \langle
p_n^2 /m \rangle$ where the brackets designate a statistical
average. Expressing $d H_n / dt$ with the Hamilton equations and using
the continuity equation from the energy flux,
\begin{equation}
  \label{eq:continuity}
  \dfrac{dH(x,t)}{dt} + \dfrac{\partial j(x,t)}{\partial x} = 0,
\end{equation}
in a finite difference form leads to a discrete expression for the local
heat flux:
\begin{equation}
  \label{eq:flux}
  j_n = K \; \langle \dot{y}_n (y_{n+1} - y_{n}) \rangle \; .
\end{equation}
The thermal properties of the model subjected to a temperature
difference, in a steady state, can be studied by molecular dynamics
simulations by imposing fixed temperatures $T_1$ and $T_2$ at the two
ends with Langevin thermostats. The simulations have to be carried long
enough to reach a steady state in which the heat flux $j$ is constant
along the lattice.

\begin{figure}[!h]
\centerline{\includegraphics*[width=0.55\textwidth]{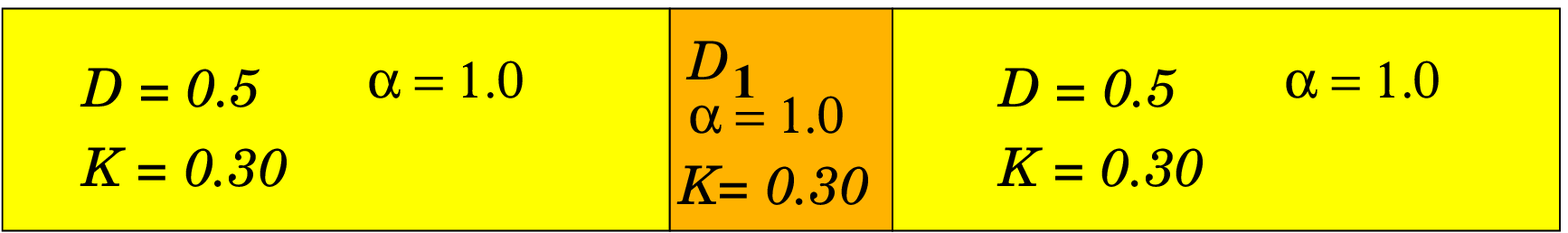}}
\vspace{0.2cm}
\centerline{\includegraphics*[width=0.55\textwidth]{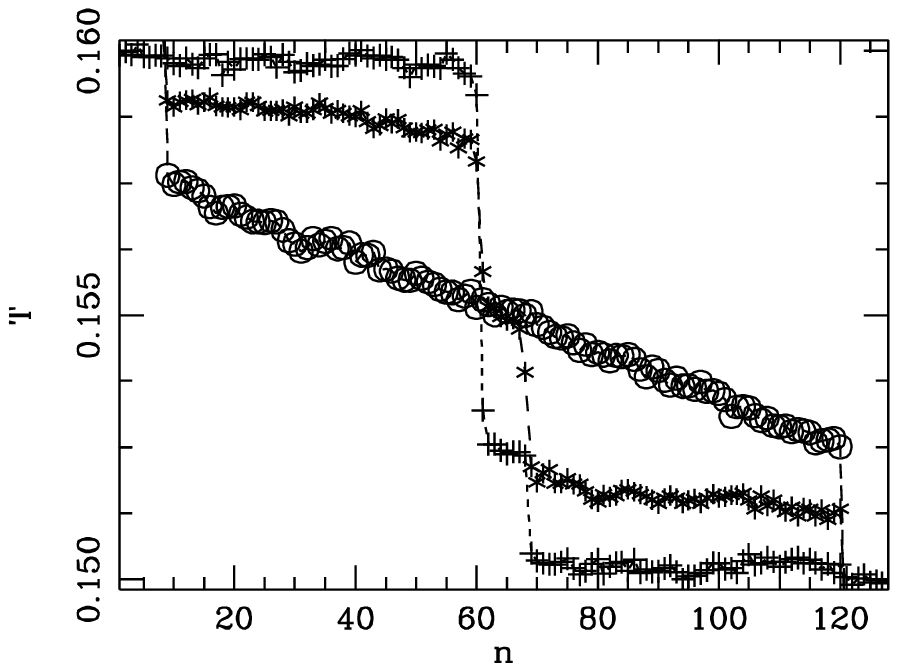}}
\caption{Top: Model parameters in the case of an inhomogeneous lattice.
Bottom: Variation of the local temperature 
along a lattice of 128
particles described by Hamiltonian~(\ref{PBmodel}) in contact with
thermostats at temperatures $T_1 = 0.16$, $T_2 = 0.15$, in energy
units, applied to the first and last 8 particles. Results for
different values of the parameter $D_1$ : $D_1 = 0.5$ (case
of a homogeneous lattice: circles), $D_1 =
0.8$ (stars), and $D_1 = 1.2$ (crosses). }
\label{fig:inhomogeneous}
\end{figure}

If the system is homogeneous ($D_n = D$ and $\alpha_n = \alpha$ for all
$n$) such calculations show that, as expected, a well defined uniform
thermal gradient is observed along the lattice, except in the immediate
vicinity of the thermostats where a sharp temperature change is observed
due to a Kapiza resistance between the thermostats and the bulk lattice
(Fig.~\ref{fig:inhomogeneous}, circles). For large $N$ the effect of the
contact 
resistance becomes negligible. The calculation 
shows that, with a fixed temperature
difference the flux decreases as $1/N$, where $N$ is the number of
lattice sites, which indicates that the model
has a well defined thermal conductivity per unit length \cite{PeyrardEPL2006}.

If the system is inhomogeneous, by including a central region in which
the parameters are different from those in the two side domains, as shown in
Fig.~\ref{fig:inhomogeneous} (top), the
heat flow is determined by the overlap of the phonon bands in the
different regions. For the example shown in
Fig.~\ref{fig:inhomogeneous}, with $T_1=0.16$ and $T_2=0.15$, the flux
is equal to $j = 0.35\times 10^{-3}$
for $D_1 = 0.5$ (corresponding to an homogeneous lattice) and decreases
to $j=0.18 \times 10^{-3}$ for $D_1 = 0.8$ for which the phonon bands
partly overlap and $j=0.48\times 10^{-5}$ when there is no overlap between
the phonon bands.


This provides a clue on a possible way to get the
temperature dependent thermal conductivity needed to build a thermal
rectifier as shown in Sec.~\ref{subsec:diode}: a nonlinearity of the
on-site potential amounts to having temperature dependent phonon bands.
In the case of the model with Hamiltonian~(\ref{PBmodel}), this can
easily be checked by a self-consistent phonon approximation
\cite{PBD}. The idea is to expand the free energy by separating the mean
value of $y_n$, $\eta = \langle y_n \rangle$, and the deviations $u_n$
around this value $y_n = \eta + u_n$. Then the Hamiltonian is
approximated by $\mathcal{H} = \mathcal{H}_0 + \mathcal{H}_1$, where
\begin{equation}
  \mathcal{H}_0 = \sum_n \left[\dfrac{1}{2} m \dot{u}_n^2 + \dfrac{1}{2} \phi
    (u_n - u_{n-1})^2 
+  \dfrac{1}{2} \Omega_2 u_n^2\right]
\end{equation}
describes an effective harmonic lattice. The free energy can be expanded
as ${\cal F} = {\cal F}_0 +  {\cal F}_1$, where
${\cal F}_1 = \langle  \mathcal{H}_1 \rangle_0$. Then, by minimizing
${\cal F}_1$ with the variational parameters $\eta$, $\phi = K$,
$\Omega_2$, one gets the lower bound of
the effective phonon band of the lattice as
$\Omega_2 = 2 \alpha^2 D \exp[ - 2 \alpha \eta / 3 ]$. As $T$ grows so
does $\eta$, so that the band shifts downwards. Therefore, if the
central region has a value $D_1 > D$ so that the phonon bands do not
overlap at low temperature, as $T$ increases the decay of the
effective lower bound of the phonon band leads to
an increased overlap, and therefore an increased thermal
conductivity. 

\begin{figure}[!h]
\centerline{\includegraphics*[width=0.45\textwidth]{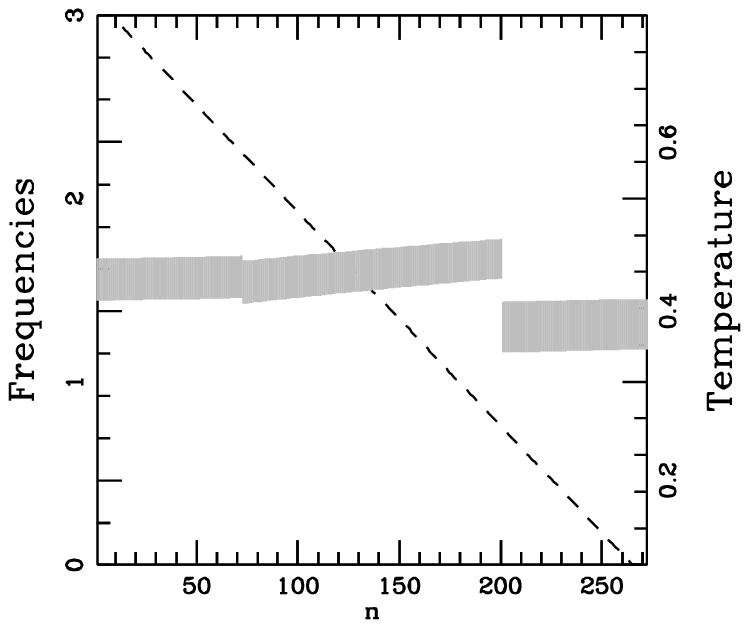}
\hspace{0.2cm}\includegraphics*[width=0.45\textwidth]{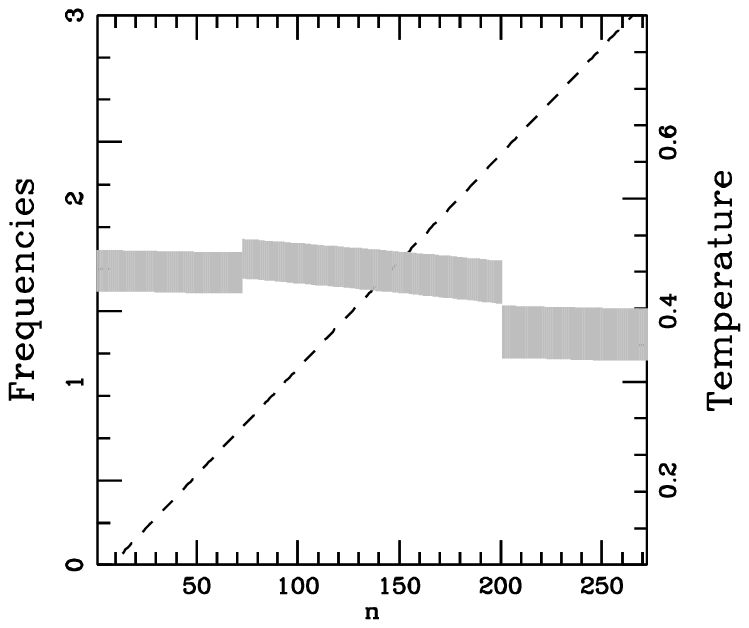}}
\vspace{0.2cm}
\centerline{\includegraphics*[width=0.45\textwidth]{freqy1091.eps}
\hspace{0.2cm}\includegraphics*[width=0.45\textwidth]{freqy1092.eps}}
\vspace{0.2cm}
\centerline{\includegraphics*[width=0.45\textwidth]{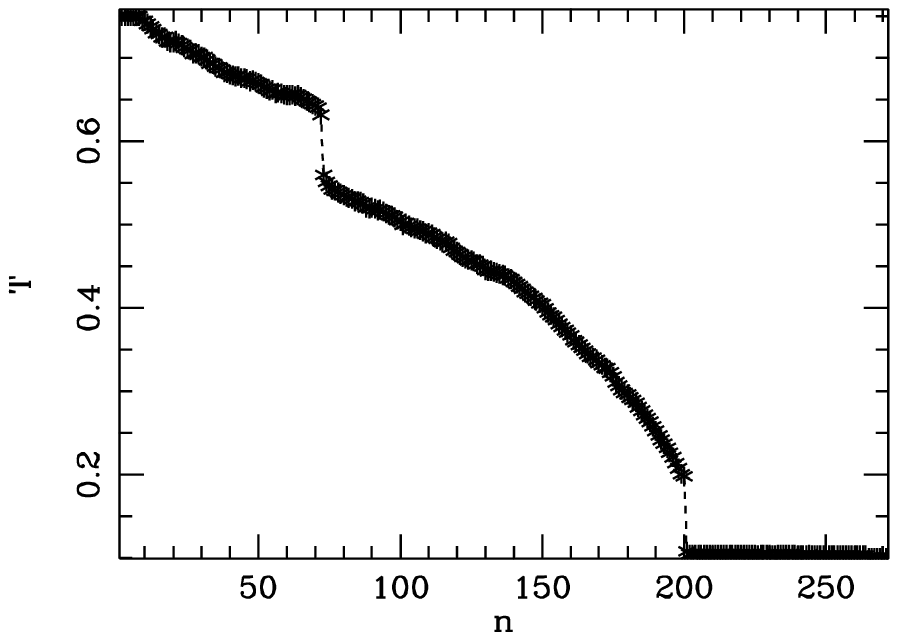}
\hspace{0.2cm}\includegraphics*[width=0.45\textwidth]{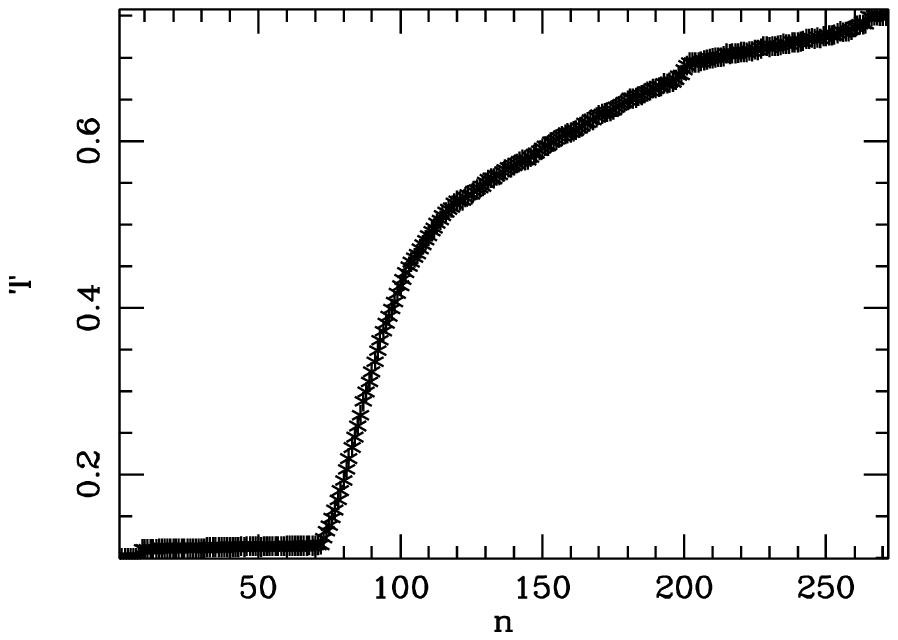}}
\caption{Properties of a model for a thermal rectifier with two
  different boundary conditions: left figures: energy flow from left to
  right $T_{\mathrm{left}} = 0.7$, $T_{\mathrm{right}} = 0.1$, right
  figures: energy flow from right to left
   $T_{\mathrm{right}} = 0.7$, $T_{\mathrm{left}} = 0.1$. The upper
   figures show the theoretical phonon bands along the device obtained
   from the self-consistent phonon approximation with the assumed
   temperature distribution shown by the dash line. The middle figures
   show the actual distribution of the phonon frequencies deduced from
   numerical simulations, and the lower figures show the variation of
   the local temperature along the system, determined from the numerical
   results. The ratio of the flux in the two directions is
   $|j_{\mathrm{right}\to\mathrm{left}}| /
   j_{\mathrm{left}\to\mathrm{right}} = 2.4$.
}
\label{fig:rectif}
\end{figure}

Expanding on these ideas one can build a thermal rectifier by
introducing the necessary asymmetry pointed out in
Sec.~\ref{subsec:diode}. Using left and right side regions with a
weak nonlinearity ($\alpha = 0.5$) and different values of the
parameter $D$ ($D_{\mathrm{left}} = 4.5$ and $D_{\mathrm{right}} = 2.8$)
and a harmonic coupling constant $K = 0.18$, one gets two domains with
phonon bands that do not overlap. Nevertheless a good thermal
conductivity can be restored with a central region with a high
nonlinearity ($\alpha = 1.1$) and $D_{\mathrm{center}} = 1.1338$ which
is such that, when the highest temperature is on the right side of the
device the variation versus space of the effective phonon band in the
central region provides a match between the left and right phonon bands,
while, if the highest temperature is on the left side of the
device, the variation versus space of the effective phonon band in the
central region leads to a large phonon-band mismatch, as shown in
Fig.~\ref{fig:rectif}. Figure~\ref{fig:rectif} shows that such a system
does indeed lead to thermal rectification because, when the hot side is
on the left, the mismatch of the phonon bands leads to temperature jumps
at the junctions between the different parts. This is due to a large
contact thermal resistance. When the hot side is on the right the
temperature evolves continuously along the device. The contact
resistances are then low, and the energy flux is 2.4 times larger in this
configuration. The calculation of the theoretical phonon bands, based on
the self-consistent phonon approximation, is only approximately correct,
first because the method itself is only approximate but also because the
calculation is made by assuming a linear temperature variation inside
the device, which is a crude approximation. However this method provides
a first step to design a rectifier, which has to be improved with the
results of the numerical simulations.

\begin{figure}[!h]
\centerline{\includegraphics*[width=0.52\textwidth]{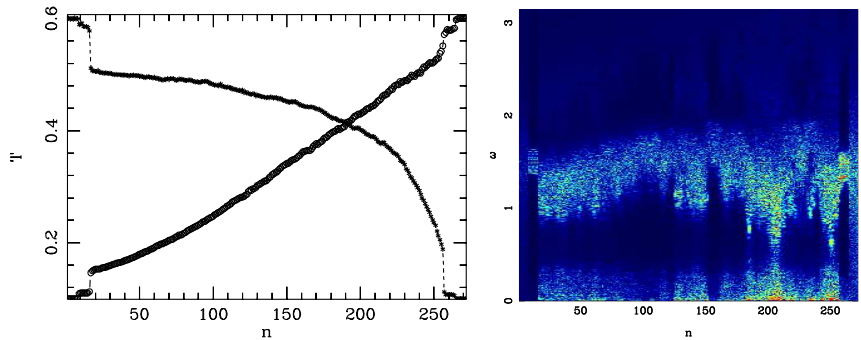}
\hspace{0.2cm}\includegraphics*[width=0.42\textwidth]{freqy1102.eps}}
  \caption{Thermal rectifier using a continuous variation of the local
    properties. Left: Temperature variation versus space inside the device
  in the two possible orientations of the thermal gradient. Right: Observed
variation of the local phonon spectra along the device when the hot side
is on the right.}
  \label{fig:continuous}
\end{figure}

The results shown in Fig.~\ref{fig:rectif} only provide a simple
illustration of what can be done with the idea of phonon-band matching,
combined to nonlinearity to allow the local phonon frequency spectra to
vary with temperature. One can imagine many possible improvements, for
instance by stacking devices, or increasing the number of interfaces, to
increase the rectifying coefficient. Another approach is to design a
system with a continuous variation of the vibrational properties
versus space, which amounts to stacking an infinity of interfaces which
have temperature dependent properties and therefore have different
transmissivities when the direction of the temperature gradient is
reversed. This allows a better control of the rectifying effect. Figure
\ref{fig:continuous} shows such an example, which has a rectifying
coefficient $R = 4.95$ and exhibits an effective phonon band which is
almost flat when the thermal gradient is in the favorable direction.

In spite of these achievements there are still several problems which
are difficult to overcome. In particular in the model thermal
rectifiers described above, 
the rectification power is small and rapidly decays to zero as the
system size increases.  A possible way to overcome this difficulty has been
discussed in \cite{CPS} where, by considering one dimensional anharmonic
chains of oscillators, empirical evidence is provided that \emph{graded mass
distribution} and \emph{long range interparticle interactions}, lead to a
substantial improvement of the thermal rectification phenomenon which moreover
does not decay to zero with increasing system size.

The system is a one-dimensional chain of $N$ oscillators described by the
Hamiltonian
\begin{equation}\label{Hamiltonian}
\mathcal{H}=\sum_{j=1}^{N}\left(\frac{p_j^2}{2m_j}+ \frac{q_{j}^{4}}{4}\right)
+\sum_{i, j}\frac{\left(q_j - q_{i}\right)^{2}}{2 + 2|i-j|^{\lambda}}~,
\end{equation} 
where $q_j$ is the displacement of the $j$th particle with mass $m_j$ and
momentum $p_j$ from its equilibrium position.  A graded mass distribution is
used.  The exponent $\lambda$ controls the decay of the interparticle
interactions with distance.

In view of previous results \cite{baowen2007,WPC}, 
it is expected that in a system with
graded mass distribution, e.g.  $m_{1}<m_{2}< \ldots <m_{N}$, thermal
rectification will be present, even for the simple case of nearest neighbor
interaction (NN). Long range interactions (LRI) introduce new channels for the
heat transport through the
new links (interactions) between the different sites. Moreover
in a graded system, the new channels connect distant particles with very
different masses. Therefore new, asymmetric channels, are created which in
turn favors the asymmetric flow, {\it i.e.}, rectification. Hence, by
introducing long range interactions in a graded system, an increase of the
thermal rectification is expected. Moreover, as we increase the system size,
new particles are introduced that, in the case of long range interactions,
create new channels for the heat current. This may avoid the usual decay of
rectification with increasing system length.

\begin{figure}[!h]
\centerline{\includegraphics*[width=0.55\textwidth]{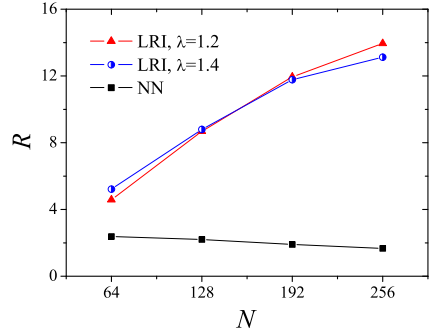}}
\caption{Dependence of rectification factor on the system size $N$. Here $T_1=9.5$, $T_2=0.5$, $m_{1}=1$. Triangles are for LRI with $\lambda = 1.2$, circles are for LRI with $\lambda = 1.4$, squares are for the NN case. The mass gradient is fixed ($m_N=2$ for N=64; $m_N=3$ for N=128; $m_N=4$ for N=192; $m_N =5$ for N=256.).}
\label{fig:longrange}
\end{figure}

In Fig.~\ref{fig:longrange} we plot the rectification factor as a function of
the system size. Here the mass gradient is fixed.  It is seen that the
presence of LRI leads to a very large rectification and prevents the decay of
the rectification factor with the system size.  Strictly speaking we cannot
make any claim for larger system sizes. However it is clear from
Fig.~\ref{fig:longrange} that the N-dependence for the LRI case is
qualitatively different from the NN case where the decay of the rectification
factor with $N$ is observed.  


\subsection{Model in higher dimension}
\label{subsec:deuxd}

Extending the same concept to higher dimension is of course important
for actual applications. For instance two dimensional models could
describe smart conducting layers to carry heat out of some nano-devices.
The same idea of playing with the phonon bands is indeed also valid in
two dimensions.

\begin{figure}[!h]
\centerline{\includegraphics*[width=0.6\textwidth]{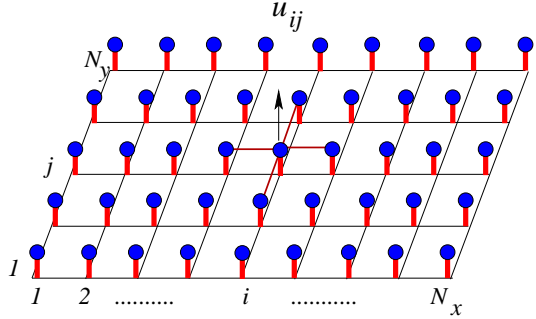}}
  \caption{Two dimensional lattice described by Hamiltonian (\ref{eq:hamil2d}).}
  \label{fig:lattice2d}
\end{figure}

Figure~\ref{fig:lattice2d} shows a two-dimensional lattice of local
oscillators which is described by the Hamiltonian
\begin{align}
\mathcal{H} = \sum_{i=1,N_x,j=1,N_y}  & \left[\dfrac{p_{ij}^2}{2m}
+ \dfrac{1}{2} C_x(i,j) \big(u_{i+1,j} - u_{i,j}\big)^2
+\dfrac{1}{2} C_x(i-1,j) \big(u_{i,j} - u_{i-1,j}\big)^2
\right.
\nonumber\\
&+ \dfrac{1}{2} C_y(i,j) \big(u_{i,j+1} - u_{i,j}\big)^2
+\dfrac{1}{2} C_y(i,j-1) \big(u_{i,j} - u_{i,j-1}\big)^2
\nonumber\\
&\left. +D(i,j) \left(\exp[ - \alpha(i,j) u_{i,j} ] - 1 \right)^2\right]
\label{eq:hamil2d}
\end{align}

\begin{figure}[!h]
\centerline{\includegraphics*[width=0.45\textwidth]{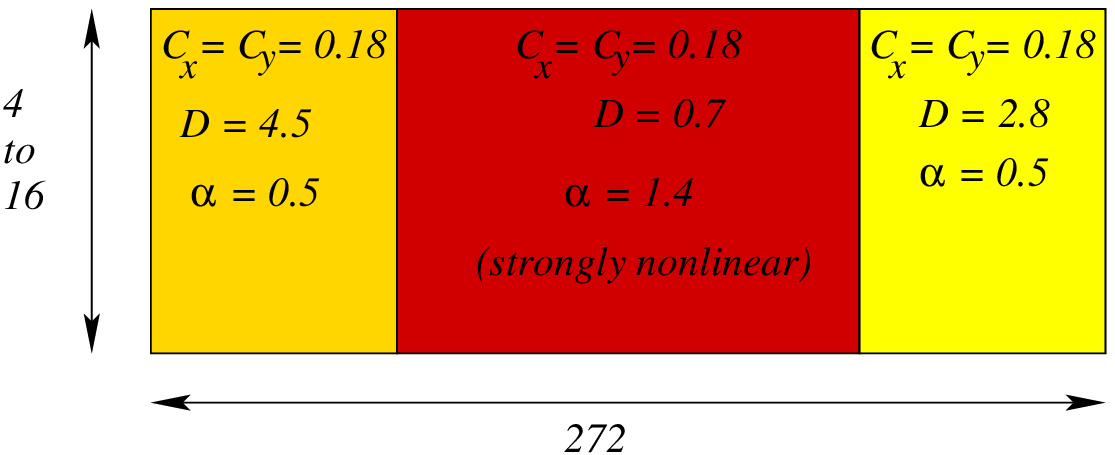}}
\vspace{0.2cm}
\centerline{\includegraphics*[width=0.45\textwidth]{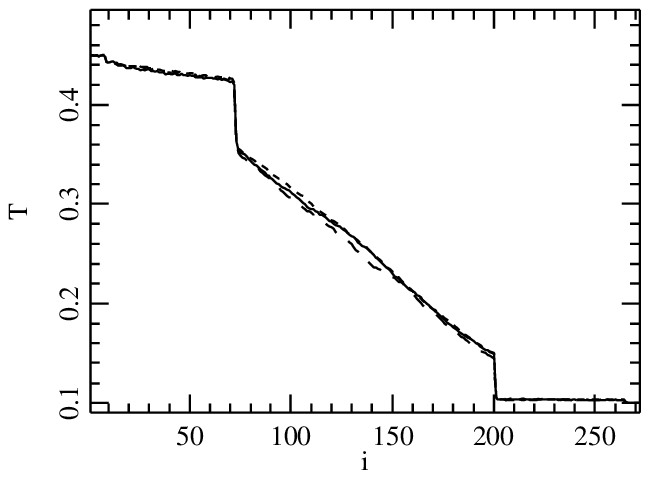}
\hspace{0.2cm}\includegraphics*[width=0.45\textwidth]{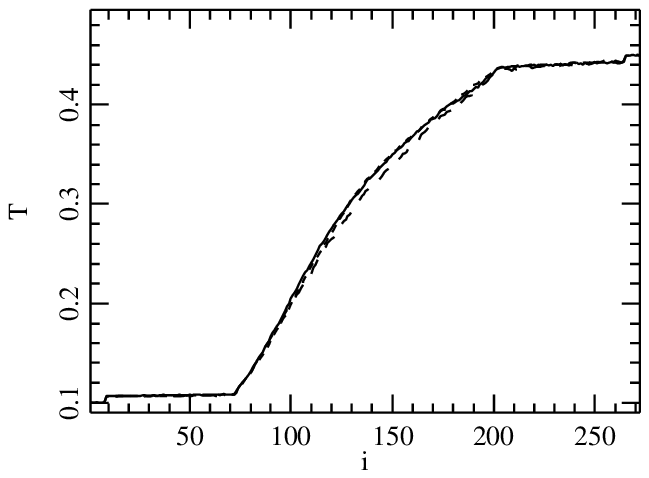}}
  \caption{Top: Schematic plot of a two-dimensional model for a thermal
    rectifier indicating the values of the parameters of
    Hamiltonian~(\ref{eq:hamil2d}) in the different regions. Bottom:
  Temperature profiles along the $x$ axis of the device for two
  opposite thermal gradients. The various lines (continuous and dash
  lines) correspond to different lattice sizes in the $y$ direction
  (from 4 to 16).}
  \label{fig:diode2d}
\end{figure}

With appropriate parameters, as indicated in Fig.~\ref{fig:diode2d} (top),
this system can operate as a rectifier with a rectifying ratio $R=1.69$
because, while the interfaces show a large thermal resistance causing a
sharp temperature drop if the left edge of the lattice is connected to
the hot bath, when the gradient is reversed, the temperature varies
smoothly along the device because there are no interfacial thermal
resistances (see Fig.~\ref{fig:diode2d} (bottom left and right)).

\subsection{Building an actual thermal rectifier}
\label{subsec:actualrectifs}

Actually the experimental observation of heat flow rectification has a
long history \cite{MARUCHA-1975}. The early observations made in 1975
with a
$GaAs$ crystal found a small rectification effect ($R \approx 1$), which
was strongly dependent on the location of the contacts on
the sample, but an asymmetry of the heat flow was nevertheless
clear. A first analysis made by assuming that the thermal conductivity
was the sum of a space-dependent term and a temperature dependent term
showed that the observation was compatible with the Fourier law
\cite{MARUCHA-1976}, and
provided the first view of the ideas presented in
Sec.~\ref{subsec:diode}. These results were followed by some debates
over the actual origin of the observed rectification \cite{BALCEREK},
and a rectification coefficient $R \approx 1.35$ could then be obtained
with a two-component sample made of tin in contact with $\alpha$-brass.
Heat flow rectification could also be observed with a carbon nanotube
\cite{CHANG2006},
loaded with $C_9H_{16}Pt$  molecules on one part of its length, but the
origin of the rectification was still not clearly established.

More recently the ideas presented in Sec.~\ref{subsec:diode}
were systematically
exploited to build rectifiers \cite{TERASAKI}, using two cobalt oxides
with different thermal conductivities. The vicinity of a structural
phase transition could be used to enhance the temperature dependence of
the thermal conductivity \cite{Kobayashi} and the asymmetry of the shape
has been exploited to vary the spatial dependence of $\kappa(x,T)$
\cite{Sawaki}. The measurements show a good quantitative agreement with
the results of Sec.~\ref{subsec:diode} and \cite{PeyrardEPL2006},
if one takes into account the experimental data for the thermal
conductivity of the materials used in the device. A quantitative
microscopic calculation of $\kappa (T)$ is however a harder
challenge. The control of the temperature dependence of $\kappa$,
trough a control of the phonon bands, discussed in
Sec.~\ref{subsec:oned} is only one possibility but other mechanisms
can be considered whether they use a structural change through a phase
transition, or variations of the density of mobile carriers in materials
which are also electrical conductors. And of course,
in such materials the use of an
electric field to manipulate the spatial distribution of the carriers
in a solid state device can also open other possibilities to control the
heat flow.
It is also worth mentioning recent experimental 
implementations of thermal rectifiers, exploiting 
phononic~\cite{Tian2012}, 
electronic~\cite{Giazotto2014}, or 
photonic~\cite{Dames} thermal currents.
Possibilities to manipulate phonons and 
devise heat diodes, transistors, thermal logic gates
and thermal memories are reviewed in Ref.~\cite{baowen2012}.


\section{Thermoelectric Efficiency}
\label{sec:TE}

Thermoelectricity concerns  the conversion of  temperature differences
into  electric potential  or vice-versa.  It  can be  used to  perform
useful electrical  work or to pump  heat from a cold to a hot place, thus
performing  refrigeration. Although  thermoelectricity  was discovered
about 200  years ago, a  strong interest of the  scientific community
arose  only in  the  1950's  when Abram  Ioffe  discovered that  doped
semiconductors  exhibit relatively  large thermoelectric  effect. This
initiated an intense research activity in semiconductors physics which
was not motivated by microelectronics but by Ioffe's suggestion that
home    refrigerators    could    be   built    with    semiconductors
\cite{Mahan1997,Majumdar2004}.  
As  a  result  of  these  efforts  the
thermoelectric material $\mathrm{Bi}_2\mathrm{Te}_3$ was developed for
commercial purposes. However this activity lasted only few years until
the mid 1960's since, in spite of all efforts and consideration of all
type of  semiconductors, it turned out  that thermoelectric refrigerators  
have  still poor  efficiency  as  compared  to compressor  based
refrigerators.  Nowadays  Peltier  refrigerators  are mainly  used  in
situations in which reliability and  quiet operation, and not the cost
and  conversion efficiency, is  the main  concern, like  equipments in
medical applications, space  probes, etc.

In the last two decades  thermoelectricity has experienced a 
renewed interest~\cite{Cahill2003,Dresselhaus2007,Bell2008,Boukai2008,Snyder2008,Shakouri2011,Dubi2011,TEbook}
due to the perspectives  of using tailored thermoelectric nanomaterials,
where a dramatic enhancement of the energy harvesting performances 
can be envisaged~\cite{hicks}.   
Indeed layering in low-dimensional systems may reduce the
phonon thermal conductivity as phonons can be scattered
by the interfaces between layers. Moreover, sharp features
in the electronic density of states, favorable for
thermoelectric conversion (see the discussion below)
are in principle possible due to quantum confinement.
Recent  efforts have focused on  one  hand, on the study
of nanostructured materials
and on  the other  hand, in
understanding the  fundamental dynamical mechanisms  which control the
coupled transport of heat and particles \cite{TEreview}.

\subsection{The thermoelectric figure of merit $ZT$}
\label{sec:ZT}

For a material subject to  a temperature gradient $\nabla T$ and a
external uniform electric field $\mathcal{E}$, 
within linear response the equations describing thermoelectric transport
are
\begin{equation} \label{eq:TE}
\begin{array}{lcl}
j_q  & \ = \ & - \kappa^\prime \nabla T +  \sigma \Pi \mathcal{E}, \,
\\
j_e & \ = \ & - \sigma S \nabla T + \sigma \mathcal{E}, \,
\end{array}
\end{equation}
where  $j_q$ and  $j_e$ denote  the heat  and electric local
currents  appearing  in  the  material  due to  the  external  forcing,
$\sigma$  is the  coefficient  of electrical  conductivity,  $S$ is  the
thermopower (or Seebeck coefficient), $\Pi$ is the Peltier 
coefficient, and $\kappa^\prime$ is
the heat conductivity  measured at zero electric field  and is related
to  the usual  heat conductivity  $\kappa$ measured  at  zero electric
current as $\kappa^\prime = \kappa + T \sigma S\Pi$. From \eref{eq:TE}
the  usual  phenomenological  relations  follow:  if  the  temperature
gradient vanishes, $\nabla T=0$,  then $j_e = \sigma {\cal E}$
is Ohm's law and the Peltier coefficient $\Pi=j_q/j_e$.  
If  the electric current vanishes,  $j_e =
0$,  then $j_q  = -\kappa  \nabla  T$ is  
Fourier's law, and ${\cal E}  = S  \nabla T$,  which is  the definition  of the
thermopower.  
We start by considering systems with time-reversal symmetry,
for which the Onsager reciprocity relations 
imply $\Pi = T S$ (see Sect.~\ref{sec:onsager}).

\begin{figure}[!h]
\centerline{\includegraphics*[width=0.55\textwidth]{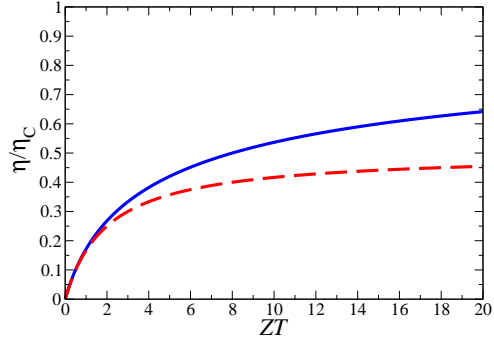}}
\caption{Linear response efficiency for heat to work conversion,
in units of the Carnot efficiency $\eta_C$, as a function of the figure
of merit $ZT$. The top and the bottom curve correspond to the
maximum efficiency $\eta_{\rm max}$ and to the efficiency at
the maximum power $\eta(P_{\rm max})$, respectively.}
\label{fig:ZT}
\end{figure}

The suitability of a  thermoelectric material for energy conversion or
electronic   refrigeration   is   evaluated   by   the   dimensionless
thermoelectric figure of merit $ZT$ \cite{Ioffe}
\begin{equation} \label{eq:ZT-def}
ZT = \frac{\sigma S^2}{\kappa}\,T \ ,
\end{equation}
as follows.
Consider a material maintained on  one end at temperature $T_\HH$ and
on  the other  at  temperature  $T_\CC$, and  subject  to an  external
electric field $\mathcal{E}$. Then $ZT$ is related to the efficiency 
$\eta=P/j_q$
of converting  the heat  current $j_q$ (flowing between  the
thermal baths) into  electric
power  $P\equiv\mathcal{E} j_e$, generated  by attaching  the 
thermoelectric  element  to an Ohmic impedance. 
If we optimize the efficiency over $\mathcal{E}$ we obtain
the \emph{maximum efficiency}
\begin{equation} \label{eq:efficiency}
\eta_{\rm max} = \eta_\mathrm{C} \, \frac{\sqrt{ZT + 1} -
  1}{\sqrt{ZT + 1} + 1} \ ,
\end{equation}
where  $\eta_\mathrm{C}=1-T_\CC/T_\HH$ is  the  Carnot efficiency
and $T  = (T_\HH + T_\CC)/2$ is the average temperature.
Thermodynamics only imposes (see Sect.~\ref{sec:onsager})
$ZT\ge 0$ and $\eta_{\rm max}$ is a monotonous growing function of $ZT$
(see Fig.~\ref{fig:ZT}), 
with  $\eta_{\rm max}=0$ when $ZT=0$ and   
$\eta_{\rm max}\to\eta_\mathrm{C}$ when $ZT\to\infty$.  

The Carnot efficiency is obtained for reversible quasi-static transformations,
which require infinite time and consequently the extracted power
is zero. An important question is how much
the efficiency deteriorates when transformations are operated in a
finite time. This is a central question in the field
of \emph{finite-time thermodynamics} \cite{Andresen11}.
Hence, the notion of \emph{efficiency at maximum power} 
$\eta(P_{\rm max})$ was introduced: it is obtained
by optimizing over  $\mathcal{E}$ the power $P$ rather than the 
efficiency $\eta$. Within linear response we obtain \cite{vandenbroeck}
\begin{equation}\label{eq:effmaxpower}
\eta(P_{\rm max})=
\frac{\eta_{C}}{2}\frac{ZT}{ZT+2}\ .
\end{equation}
Note that also $\eta(P_{\rm max})$ is a growing function of $ZT$ 
(see Fig.~\ref{fig:ZT}).
In the limit $ZT \rightarrow \infty$, $\eta(P_{\rm max})$ takes 
its maximum value of $\eta_{C}/2$. Such value also corresponds to the
linear response expansion of the so-called 
Curzon-Ahlborn upper bound
\cite{Yvon1955,Chambadal1957,Novikov1958,Curzon1975,vandenbroeck,Schmiedl2008,Esposito2010,Goupil2012}. 
Therefore, high values  of $ZT$ are favorable for thermoelectric conversion.

Nowadays,  most  efficient thermoelectric  devices  operate at  around
$ZT\approx1$, whilst it is generally accepted that $ZT> 3$ is the
target  value  for  efficient, commercially  competing,  thermoelectric
technology  \cite{Majumdar2004}.  The  great challenge  to increase
thermoelectric efficiency relies on understanding the microscopic mechanisms that
may allow to control individually $S$, $\sigma$ and $\kappa$.  However,
the  different  transport  coefficients are  generally  interdependent
making  optimisation extremely difficult  and so  far, no  clear paths
exist which  may lead to reach  that target.  A  particular example of
this interdependence is the Wiedemann-Franz law \cite{ashcroft1976} which states that for
metallic materials,  $\sigma$ and $\kappa$  are, as a matter  of fact,
proportional,  thus  making metals  poor  thermoelectric materials  in
general.

Note that $ZT$ is related  to the heat conductivities defined above as
$\frac{\kappa^\prime}{\kappa}=1+ZT$, which has been used in
Ref.~\cite{Vining1997} to make an analogy between a classical 
heat engine 
and a thermoelectric material. The used correspondence is 
$N\to V$ and $\mu_e\to -p$, with $N$ number of charge carriers,
$\mu_e$ the electrochemical potential, and
$V,p$ volume and pressure of the gas in the engine. As a consequence,
$\frac{\kappa^\prime}{\kappa}\to \frac{c_p}{c_V}$, with $c_p$ and $c_V$ specific
heat at constant pressure and volume, respectively. The ratio $\frac{c_p}{c_V}$ is 
bounded for ideal (non interacting) gases, but diverges at the gas-liquid critical
conditions. These considerations suggest that large values of $ZT$
could be expected near electronic phase transitions, for systems 
with strong interactions between the charge carriers \cite{Ouerdane2014}.

 \subsection{The Onsager matrix}
\label{sec:onsager}

Let us consider  a system of particles enclosed  in a chamber, coupled
to  two  particle reservoirs.   Calling  the  energy  balance for  the
thermoelectric process, the energy current  can be written in terms of
the   heat   and  electric   currents   as   $j_u   =  j_q   +
\frac{\mu_e}{e}\,j_e$,  where  $\mu_e$  is  the  electrochemical
potential.    For   particles   having   electric   charge   $e$   the
electrochemical  potential is  simply  $\mu_e =  e\phi$, where  $\phi$
is the  ordinary electrostatic  potential  ($\mathcal{E} =  -\nabla
\phi$).  Assuming that  the particles are the only  carriers of heat,
one may  interchange the  electrochemical potential with  the chemical
potential $\mu$ corresponding to the work generated by the exchange of
particles between the system  and the reservoirs. 
Within the linear response regime, the energy current and the 
particle current $j_\rho=\frac{1}{e}\,j_e$ are related to
the  conjugated thermodynamic forces  (gradient of  chemical potential
$\mu$ and gradient of temperature $T$) as \cite{Degroot1962,callen}
\begin{equation} \label{lineareqs}
{\bf j}=\mathbb{L}\,{\bf F}\ ,
\end{equation}
where
${\bf j}\equiv(j_\rho,j_u)^t$, ${\bf F}\equiv(\nabla(-\mu/T),\nabla(1/T))^t$, and 
\begin{equation} \label{onsager}
\mathbb{L} \equiv \left(
\begin{array}{cc}
L_{\rho \rho} & L_{\rho u} \\
L_{u \rho} & L_{u u}
\end{array}
\right) 
\end{equation}
is  the  Onsager matrix  of  kinetic  transport  coefficients. In  the
absence   of  magnetic   fields  (or   other  effects   breaking  time
reversibility),  the  Onsager reciprocity  relations  state that  the
crossed  kinetic   coefficients  are  equal:   $L_{\rho  u}=L_{u\rho}$.
Moreover, the second law  of thermodynamics imposes that the entropy 
production rate 
$\dot{s}={\bf j}\cdot {\bf F}=j_\rho \nabla(-\mu/T)+
j_u \nabla(1/T)\ge 0$. 
Therefore $\mathbb{L}$ has to be nonnegative,
i.e. $L_{\rho\rho},L_{uu}\ge 0$ and $\det \mathbb{L}\ge 0$.


The kinetic  coefficients $L_{ij}$ 
are related to
the thermoelectric transport coefficients as
\begin{equation} \label{transport}
\sigma=\frac{e^2}{T}L_{\rho\rho} \ ,
\quad\kappa=\frac{1}{T^2}\frac{\det \mathbb{L}}{L_{\rho\rho}},
\quad S=\frac{1}{eT}\left(\frac{L_{\rho u}}{L_{\rho\rho}}-\mu\right)=\frac{\Pi}{T} \ ,
\end{equation}
where the temperature $T$
and chemical potential $\mu$ are taken as mean values in the bulk.
Moreover,   using    Eqs.~(\ref{eq:ZT-def})   and
(\ref{transport}), the thermoelectric figure of merit reads
\begin{equation} \label{ZT}
ZT=
\frac{(L_{u\rho}-\mu L_{\rho\rho})^2}{\det \mathbb{L}}
\ .
\end{equation}
Note that the limit  $ZT\to\infty$ can be reached if and only if the
Onsager matrix $\mathbb{L}$ is ill-conditioned, namely when the ratio
\begin{equation}\label{eq:cond}
\frac{\left[\tr (\mathbb{L})\right]^2}{\det\mathbb{L}} \rightarrow \infty
\end{equation}
and therefore the linear system (\ref{lineareqs}) becomes
singular.
That is, the Carnot efficiency is obtained when the energy current and
the particle current become proportional: $j_u=c j_\rho$, with the
proportionality factor $c$ independent of the values of the 
applied thermodynamic forces. Such condition is refereed to as 
\emph{tight coupling} condition.

\subsection{Non-interacting systems}
\label{sec:nonint}

We consider a system whose ends are in contact with
left/right baths (reservoirs), which are able to exchange energy and particles
with the system, at fixed temperature $T_\alpha$
and chemical potential $\mu_\alpha$, where $\alpha=L,R$
denotes the left/right bath. The reservoirs are modeled as 
infinite ideal gases, and therefore particle velocities 
are described by the Maxwell-Boltzmann distribution.
We use a stochastic model of the thermochemical
baths~\cite{M-M2001,Larralde2003}: Whenever a particle of the system crosses
the boundary which separates the system from the left or right reservoir,
it is removed. On the other hand, particles are injected into the
system from the boundaries, with rates $\gamma_\alpha$ computed
by counting how many particles from reservoir $\alpha$ cross
the reservoir-system boundary per unit time.
For one-dimensional reservoirs we obtain
$\gamma_\alpha=\frac{1}{h\beta_\alpha} e^{\beta_\alpha \mu_\alpha}$,
where $\beta_\alpha=1/(k_B T_\alpha)$ ($k_B$ is the Boltzmann constant
and $h$ is the Planck's constant). 
Assuming that both energy  and charge are carried
only by non-interacting particles, like in a dilute gas, we arrive
at simple expressions for the particle and 
heat currents~\cite{Saito2010}:
\begin{equation}
j_\rho=\frac{1}{h}\int_0^\infty d\epsilon
\left( e^{-\beta_L (\epsilon-\mu_L)}-
e^{-\beta_R (\epsilon-\mu_R)} \right) {\tau}(\epsilon)\ ,
\label{eq:jrhodelta}
\end{equation}
\begin{equation}
j_{q,\alpha}=\frac{1}{h}\int_0^\infty d\epsilon (\epsilon - \mu_\alpha)
\left( e^{-\beta_L (\epsilon-\mu_L)}-
e^{-\beta_R (\epsilon-\mu_R)} \right) {\tau}(\epsilon)\ ,
\label{eq:jqalpha}
\end{equation}
where $j_{q,\alpha}$ is the heat current from reservoir $\alpha$ and
${\tau}(\epsilon)$ denotes the transmission
probability for a particle with energy $\epsilon$ to transit from
one end to the other end of the system ($0\le {\tau}(\epsilon)\le 1$).
The thermoelectric efficiency is then given by
(we assume $T_L>T_R$, $\mu_R>\mu_L$, and $j_\rho,j_{q,L}\ge 0$)
\begin{equation}
\eta=\frac{j_{q,L}-j_{q,R}}{j_{q,L}}=
\frac{(\mu_R-\mu_L)\int_0^\infty d\epsilon
\left( e^{-\beta_L (\epsilon-\mu_L)}-
e^{-\beta_R (\epsilon-\mu_R)} \right)
{\tau}(\epsilon)}{\int_0^\infty d\epsilon (\epsilon - \mu_L)
\left( e^{-\beta_L (\epsilon-\mu_L)}-
e^{-\beta_R (\epsilon-\mu_R)} \right) {\tau}(\epsilon)}\ .
\end{equation}
When the transmission is possible only within a tiny energy
window around $\epsilon=\epsilon_\star$, the efficiency reads
\begin{equation}
\eta= \frac{\mu_R-\mu_L}{\epsilon_\star-\mu_L}\ .
\label{eq:etalinke}
\end{equation}
In the limit $j_\rho\to 0$, corresponding to reversible
transport~\cite{Linke2002,Humphrey2005}, we get $\epsilon_\star$
from Eq.~(\ref{eq:jrhodelta}):
\begin{equation}
\epsilon_\star=\frac{\beta_L\mu_L-\beta_R\mu_R}{\beta_L-\beta_R}\ .
\label{eq:etalinke2}
\end{equation}
Substituting such $\epsilon_\star$ in Eq.~(\ref{eq:etalinke}),
we obtain the Carnot efficiency $\eta=\eta_C=1-T_R/T_L$.
Such delta-like energy-filtering
mechanism for increasing thermoelectric efficiency
has been pointed out in Refs.~\cite{Mahan1996,Linke2002,Humphrey2005}.
As remarked above,  Carnot  efficiency  is
obtained  in  the limit  of  zero  particle  current, corresponding  to
zero entropy  production and zero output power.
However, high  values  of  $ZT$  can  still  be  achieved  with  sharply-peaked
transmission  functions  without  greatly  reducing the  output  power
\cite{ODwyer2005,Vashaee2004}. 

In the linear response regime, using a delta-like energy filtering,
i.e. ${\tau}(\epsilon)=1$ in a tiny interval of width $\delta \epsilon$
around
some energy $\bar{\epsilon}$ and $0$ otherwise, we obtain
\begin{equation}
L_{\rho \rho} = {\Lambda(\delta\epsilon) \over hk_B}
e^{-\beta (\bar{\epsilon} -\mu )} , \;
L_{u\rho} =  L_{\rho u}=
{\Lambda \bar{\epsilon}
(\delta \epsilon)\over hk_B} e^{-\beta (\bar{\epsilon} -\mu )}  ,  \;
L_{uu} = {\Lambda \bar{\epsilon}^2 (\delta \epsilon)\over hk_B} e^{-\beta
(\bar{\epsilon} -\mu )}\  , 
\label{eq:deltalinear}
\end{equation}
where $\Lambda$ is the length of system.
From these relations we immediately derive that the Onsager matrix
is ill-conditioned and therefore $ZT=\infty$ and
$\eta=\eta_C$.
We point out that the parameters $\bar{\epsilon}$ and
$\delta \epsilon$ characterizing the transmission window,
appear in the Onsager matrix elements (\ref{eq:deltalinear})
and therefore are assumed to be
independent of the applied temperature and
chemical potential
gradients. On the other hand, the energy
$\epsilon_\star$ in Eqs.~(\ref{eq:etalinke}) and (\ref{eq:etalinke2})
depends on the applied gradients.
There is of course no contradiction since
(\ref{eq:etalinke}) and (\ref{eq:etalinke2}) have general validity
beyond the linear response regime.

A dynamical realization of the energy-filtering mechanism
was discussed in Ref.~\cite{Casati2008}. 
We start by writing for a gas of non-interacting particles 
the microscopic instantaneous charge and energy currents per particle at position 
${\bf r}^*$ and time $t$:
\begin{eqnarray} \label{eq:inst-currents}
\iota_\rho({\bf r}^*,t) & \ = \ & v_x\delta({\bf r}^*-{\bf r}(t)) \ , \\
\iota_u({\bf r}^*,t) & \ = \ & \epsilon(t) v_x({\bf r}(t),t)\delta({\bf r}^*-{\bf r}(t)) \ , 
\end{eqnarray}
where  $\epsilon$ is  the energy  of the  particle, ${\bf r}$  its position  and  $v_x$ its
velocity along the direction of the currents.  
The  thermodynamic averages  of the  two currents
become  proportional  precisely  when  the
variables $\epsilon$ and $v_x$ are uncorrelated:
\begin{equation} \label{eq:main}
j_u=\ave{\iota_u} =  \ave{\epsilon v_x}=\ave{\epsilon}\ave{v_x} = \ave{\epsilon}\ave{\iota_\rho}=
\ave{\epsilon} j_\rho\ .
\end{equation}
Therefore,  $ZT=\infty$ follows  from  the fact  that  the average  particle's
energy $\ave{\epsilon}$ does not depend  on the thermodynamic forces.  In the context
of classical physics this happens for instance in the limit of large number of
internal degrees  of freedom,  provided the dynamics is  ergodic.

\begin{figure}[!h]
  \centerline{\includegraphics*[width=0.65\textwidth]{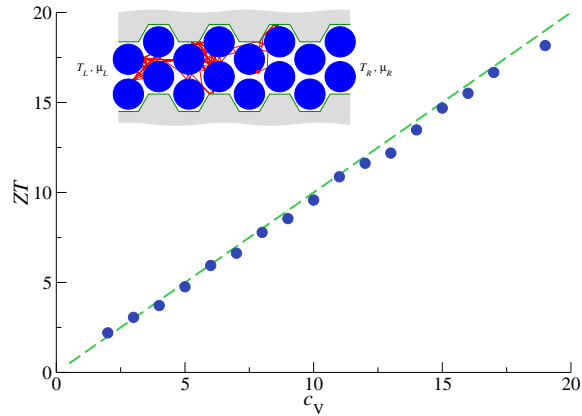}}
  \caption{Figure of  merit $ZT$  as a function  of the  heat capacity
    $c_V$, at $\mu =  0$. Numerical results are obtained 
    from nonequilibrium simulations (for
    the  details of the  simulations see  Ref.~\cite{Casati2008}). The
    dashed   line  corresponds   to  the   analytical   expression  of
    Eq.~(\ref{eq:ZTmain}). Inset: schematic drawing of the model 
    used in the numerical simulations.}
  \label{fig:TE-ZTnonint}
\end{figure}

This observation was used in Ref.~\cite{Casati2008} where an ergodic gas of
non-interacting  particles  with $d_{\rm int}$ internal  degrees  of  freedom 
in a $d-dimensional$ chamber connected to reservoirs was
studied. It was  shown  that for such systems the thermoelectric figure of merit 
becomes
\begin{equation}\label{eq:ZTmain}
ZT = \frac{1}{c_V}\left(c_V - \frac{\mu}{T}\right)^2 \ ,
\end{equation}
where $c_V =  c^*_V + 1/2$ and $c^*_V=D/2$  ($D=d+d_{\rm int}$) 
is the dimensionless heat capacity at constant volume of the gas.
Fig.~\ref{fig:TE-ZTnonint} shows the  figure of merit $ZT$ numerically
computed for  a gas  of noninteracting point-like particles  as a 
function of the specific heat
(internal degrees of freedom are modeled as free  rotating modes). The
particles evolve inside a Lorentz  gas channel with finite horizon, so
that   the    particles   motion   is   diffusive    (see   the inset   of
Fig.~\ref{fig:TE-ZTnonint}).   The  channel   is   connected  at   its
boundaries to stochastic reservoirs at different temperatures
and chemical potentials. The numerical results confirm the analytical
expression of Eq.~(\ref{eq:ZTmain}).
The  simple mechanism  for the  growth of  $ZT$ also implies that  the  equilibrium
distribution of the  particle energy per degree of  freedom becomes more
sharply peaked as $D$ increases. 

We point out that, while the discussion in this section was focused
on classical systems, noninteracting systems can be easily treated in
quantum mechanics too by means of the Landauer-B\"uttiker  formalism,
\cite{Datta1995,Imry1997}. In this approach, the 
particle and heat currents are given, similarly to 
Eqs.~(\ref{eq:jrhodelta}) and (\ref{eq:jqalpha}), 
in terms of integrals over the energy distribution
of the particles injected from the reservoirs and the 
scattering transmission probability of the system
(for the use  of this formalism
in thermoelectricity see Ref.~\cite{TEreview}).      
Implementations of the energy
filtering mechanism may be possible in, e.g., nanowires or nanostructured 
materials for which the shape of the transmission function
can be controlled more easily than in bulk materials.
Finally, we note that the results of this section are obtained in the absence
of  phonon heat leaks.

\subsection{Interacting systems}
\label{sec:int}

\begin{figure}[!h]
  \centerline{\includegraphics*[width=0.55\textwidth]{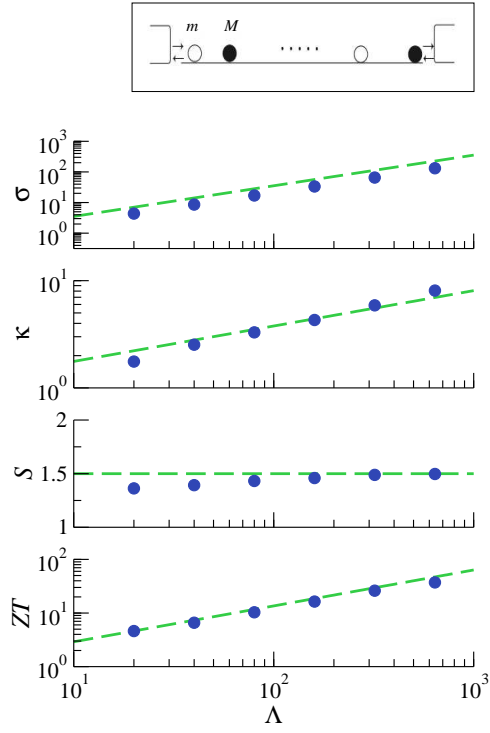}}
  \caption{Thermoelectric     transport    coefficients     for    the
    one-dimensional  di-atomic  disordered  hard-point  gas  model,   
    as  a   function  of   the   system  size
    $\Lambda$.  The  dashed curves correspond  from top to  bottom to
    $\sigma\sim\Lambda$,        $\kappa\sim\Lambda^{0.33}$,       
    $S= 1.5$,       and
    $ZT\sim \Lambda^{0.67}$.  In the  upper panel a schematic
    representation of the model is shown.}
  \label{fig:HPgas}
\end{figure}

The thermoelectric  properties of strongly interacting  systems are of
fundamental interest since their efficiency is not bounded by 
inherent limitations of non-interacting systems, such as the Wiedemann-Franz
law. Experiments  on some strongly correlated materials
such as sodium cobalt  oxides revealed unusually large thermopower
values  \cite{Terasaki1997,Wang2003},  due   in  part  to  the  strong
electron-electron  interactions \cite{Peterson2007}.  
Very little is known about the thermoelectric
properties of interacting systems: analytical results
are rare and numerical simulations are challenging.
The linear response  Kubo formalism has been used  to investigate the
thermoelectric   properties   of   one-dimensional   integrable   and
nonintegrable strongly correlated quantum lattice models
\cite{Arsenault2013,Peterson2007,Shastry2009,Zemljic2007}.
With regard to the simulation of classical dynamical models,
an   extension   of  the model discussed in Sect.~(\ref{sec:nonint}),
with   inter-particle  interactions  added  by
substituting the  Lorentz lattice with the rotating  Lorentz gas model
\cite{M-M2001,Larralde2003} was studied in Ref.~\cite{carlosAIP}.  It
was   shown   that while $ZT$ is bounded from above by its value
obtained at zero interaction, it still increases with $c_V$.
On the other hand, for a  one-dimensional  di-atomic  disordered
hard-point  gas coupled to  particle reservoirs
(see the upper panel in Fig.~\ref{fig:HPgas}  for a schematic representation 
of the  model), it  was  numerically  found 
\cite{Casati2009} that  $ZT$  diverges in  the
thermodynamic limit as  a power-law, $ZT \sim \langle N\rangle^\alpha$,  
where $\langle N\rangle$ is
the  average number  of particles  in the  system and $\alpha\approx 0.79$.
Note  that  if the
masses of all  particles are the same, the  dynamics is integrable 
and one can find analytically that $ZT$ is independent 
of $\langle N \rangle$ (in particular, $ZT=1$ when the chemical potential 
$\mu=0$).
Later Ref.~\cite{Saito2010} showed that the numerically observed 
large values of $ZT$ could not be explained in terms of the energy 
filtering mechanism. Indeed, the particle current 
at the position $x\in [0,\Lambda]$ ($\Lambda$ is the system size)
can be expressed as
$j_\rho= \int_0^\infty d\epsilon D(\epsilon)$,
where $D(\epsilon)\equiv D_L(\epsilon)-D_R(\epsilon)$
plays the role of ``transmission function'':
$D_L(\epsilon)$ is the density of particles with energy $\epsilon$
crossing $x$ and coming from the left side, while $D_R(\epsilon)$
is the density of particles with energy $\epsilon$ from the right side.
If the divergence of $ZT$ with $\Lambda$ was due to
energy filtering, then $D(\epsilon)$ would sharpen with increasing the 
system size. Conversely,
no sign of narrowing of $D(\epsilon)$ was found in Ref.~\cite{Saito2010},
As discussed below in Sect.~\ref{sec:cons}, the divergence of $ZT$ 
can be explained on the basis of a theoretical argument \cite{Benenti2013}
applicable to non-integrable systems with momentum conservation.

\subsubsection{Green-Kubo formula}
\label{sec:g-k}

While the Landauer-B\"uttiker approach cannot be applied to 
interacting systems, the linear response regime can be 
numerically investigated in equilibrium simulations by using
the  Green-Kubo formula.
This formula expresses the Onsager  kinetic coefficients in
terms   of   equilibrium   dynamic   correlation  functions   of   the
corresponding    current    at    finite   temperature    $\beta^{-1}$
\cite{Kubo1957,Mahan1990} as
\begin{equation}
L_{ij} = \lim_{\omega\to 0} {\rm Re} L_{ij} (\omega)\ ,
\end{equation}
where
\begin{equation}\label{eq:kubo}
L_{ij}(\omega)\equiv \lim_{\epsilon\to 0}
\int_0^\infty dt e^{-i(\omega-i\epsilon)t}
\lim_{\Omega \to\infty}
\frac{1}{\Omega}
\int_0^\beta d\tau\langle {J}_i(0) {J}_j (t+i\tau)\rangle,
\end{equation}
where $\langle  \; \cdot \;\rangle  = \left\{{\rm tr}\left[(\;\cdot\;)
    \exp^{-\beta    \mathcal{H}}\right]\right\}/{\rm    tr}    \left[\exp(-\beta
  \mathcal{H})\right] $ denotes the equilibrium expectation value at temperature
$T$, $\mathcal{H}$ is the system's Hamiltonian, $\Omega$  is the  system's volume,
and $J_i(t)=\int_\Omega d\vec{r} {j}_i(\vec{r},t)$ is the total 
current ($i=\rho,u$). 

Within the  framework of
Kubo linear  response approach, the real part  of $L_{ij}(\omega)$ can
be decomposed into a  singular contribution at zero frequency  and  a  regular  part
$L_{ij}^{\rm reg}(\omega)$ as
\begin{equation}
{\rm Re} L_{ij}(\omega)=
2\pi {D}_{ij}\delta(\omega)+L_{ij}^{\rm reg}(\omega) \ .
\label{eq:reLij}
\end{equation}
The  coefficient of the  singular part  defines the  generalized Drude
weights  ${D}_{ij}$  (for  $i=j=\rho$, we  have the  conventional
Drude  weight  ${D}_{\rho\rho}$).   Importantly, it  has  been
shown  that  non-zero  Drude  weights,  ${D}_{ij}\ne  0$,
are a signature of ballistic
transport~\cite{Zotos1997,Zotos2004,Garst2001,H-M2005}, namely in the
thermodynamic limit the kinetic  coefficients $L_{ij}$ diverge
linearly with the system size. Moreover, it has
been conjectured  that at finite temperature, an  integrable system is
an ideal conductor  characterised by a finite Drude  weight if at zero
temperature the Drude weight is  positive, while the system remains an
insulator if the  zero temperature Drude weight is  zero. On the other
hand  nonintegrable systems  are believed  to have  a  vanishing Drude
weight and thus, to exhibit normal transport.

\subsubsection{Conservation laws and thermoelectric efficiency}
\label{sec:cons}

The   way   in   which    the   dynamic   correlation   functions   in
Eq.~(\ref{eq:kubo})  decay,  determines   the  ballistic,  anomalous  or
diffusive character of the energy  and particle transport, and it has been
understood  that  this  decay  is directly  related  to  the
existence        of        conserved       dynamical        quantities
\cite{Zotos1997,Zotos2004}. For quantum spin chains and under suitable conditions,
it  has been  proved that systems
possessing local conservation  laws  exhibit  ballistic transport  at
finite temperature \cite{Ilievski2013}.

However, the  role that the  existence of
conserved quantities  plays on the thermoelectric  efficiency has been
considered only recently
\cite{Casati2009,Saito2010,Benenti2013,Benenti2014,Chen2015}.

The  decay of time  correlations for the currents 
can  be related  to the  existence of
conserved  quantities  by  using  Suzuki's formula
\cite{Suzuki1971}, which  generalizes and inequality  proposed  by  Mazur
\cite{Mazur1969}. 
Consider  a   system  of   size  $\Lambda$  and
Hamiltonian  $\mathcal{H}$,  with  a  set  of $M$  \emph{relevant conserved  quantities}  $Q_m$,
$m=1,\ldots  ,M$, namely  the commutators  $[\mathcal{H},Q_m]=0$. 
A constant of motion $Q_m$ is by definition relevant if 
it is not orthogonal to the currents under consideration, 
in our case $\langle J_\rho Q_m \rangle \ne 0$ and
$\langle J_u Q_m \rangle \ne 0$. 
It is assumed that the $M$ constants of motion are orthogonal,   i.e.,   
$\langle   Q_m   Q_m\rangle   =   \langle   Q_n^2   \rangle
\delta_{m,n}$ (this is always possible via a Gram-Schmid procedure).
Furthermore, we assume that the set $\{Q_m\}$ exhausts
all relevant 
conserved quantities.
Then using Suzuki's formula \cite{Suzuki1971}, 
we can express the finite-size Drude weights 
\begin{equation}
d_{ij}(\Lambda)\equiv  \frac{1}{2\Lambda}
\lim_{t\to\infty}\frac{1}{t}
\int_0^{t} dt' \langle J_i(t') J_j(0) \rangle
\end{equation}
in terms of the relevant conserved quantities:
\begin{equation}
d_{ij}(\Lambda)=\frac{1}{2\Lambda}
\sum_{m=1}^M \frac{\langle J_iQ_m \rangle \langle J_jQ_m
  \rangle}{\langle Q_m^2 \rangle} \ .
\label{eq:finitesizedrude}
\end{equation}
On the other hand, the thermodynamic Drude weights can also be expressed
in terms of time-averaged
current-current correlations as
\begin{equation}\label{eq:Drude}
{D}_{ij}=\lim_{t\to\infty}\lim_{\Lambda\to \infty}
\frac{1}{2\Lambda t}
\int_0^{t} dt' \langle J_i(t') J_j(0) \rangle \ .
\end{equation}
If the thermodynamic limit $\Lambda\to\infty$ commutes with the long-time
limit $t\to\infty$, then the thermodynamic Drude weights ${D}_{ij}$
can be obtained as
\begin{equation}
{D}_{ij}=\lim_{\Lambda\to\infty} d_{ij}(\Lambda)\ .
\label{eq:drudeinfty}
\end{equation}
Moreover, if the limit does not vanish we can conclude that the presence
of relevant conservation laws yields non-zero generalized Drude weights,
which in turn imply that transport is ballistic, $L_{ij}\sim \Lambda$.
As a consequence, the electrical conductivity is ballistic,
$\sigma\sim L_{\rho\rho}
\sim \Lambda$, while the thermopower is asymptotically size-independent,
$S\sim L_{u\rho}/L_{\rho\rho}\sim \Lambda^0$. 

We can see from Suzuki's formula that for systems with a single relevant
constant of motion ($M=1$), the ballistic contribution to $\det \mathbb{L}$
vanishes, since it is proportional to ${D}_{\rho\rho}{D}_{uu}-
{D}_{\rho u}^2$, which is zero from Eqs. (\ref{eq:finitesizedrude})
and (\ref{eq:drudeinfty}).
Hence, $\det \mathbb{L}$ grows slower than $L^2$,
and therefore the thermal conductivity $\kappa\sim \det{\mathbb{L}}/
L_{\rho\rho}$ grows sub-ballistically, $\kappa\sim L^\alpha$, with $\alpha<1$.
Since $\sigma\sim\Lambda$ and $S\sim\Lambda^0$, we can conclude 
that $ZT\sim \Lambda^{1-\alpha}$ \cite{Benenti2013}.
Hence $ZT$ diverges in the thermodynamic limit $\Lambda\to\infty$. This general
theoretical argument applies for instance to systems where momentum is the
only relevant conserved quantity. 

\begin{figure}[!h]
  \centerline{\includegraphics*[width=0.55\textwidth]{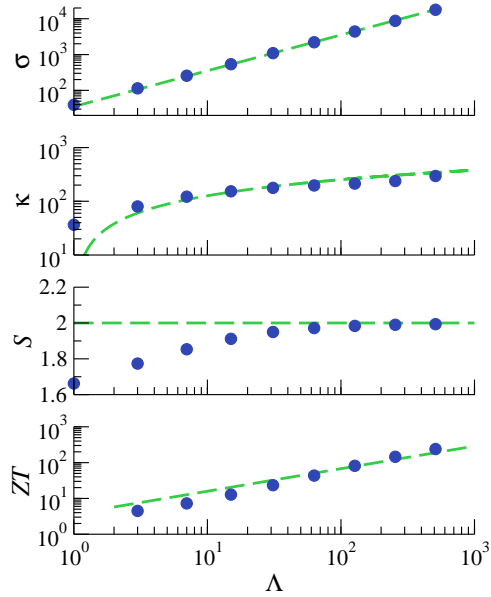}}
  \caption{Thermoelectric transport coefficients for the
    two-dimensional MPC gas of interacting particles as a function of
    the system size $\Lambda$ 
(for
    details see \cite{Benenti2014}). The dashed curves correspond from
top to bottom to $\sigma=(\pi \langle N\rangle/2m)\Lambda$ with 
$\langle N\rangle$ the mean number of
particles,  $\kappa\sim \log\Lambda$, $S=2$, and $ZT\sim\Lambda/\log\Lambda$.}
  \label{fig:kapral}
\end{figure}

It has been recently shown that these  expectations  fully  describe  the  results  obtained  for  the
one-dimensional disordered  hard-point gas, see Fig.~\ref{fig:HPgas}
and Ref.~\cite{Benenti2013}. 
This enhancement  of $ZT$ has also been verified for  more realistic
models in
Ref.~\cite{Benenti2014}, where  the  nonequilibrium steady
state properties  of  a two-dimensional  gas  of  particles interacting  through
elastic  collisions  and  enclosed  in  a box  connected  to  
reservoirs at both ends were studied numerically. 
The inter-particle collisions were modeled
by the method of Multiparticle Collision Dynamics (MPC) \cite{Kapral1999}.
Similarly to Ref.~\cite{Benenti2013}, it was found that the
generalized Drude weights are finite, leading to non decaying
current-current time correlations. As a consequence, the transport
coefficients exhibit an anomalous scaling yielding a figure
of merit that for this model diverges as $ZT \sim
\Lambda/\log\Lambda$. The logarithmic term appears as a consequence of
the existence of long time tails in the decay of the energy
current-current time correlation, typically observed in
two-dimensional hydrodynamic systems \cite{Alder1967}. The
dependence of the thermoelectric transport coefficients as a function
of the system size is shown in Fig.~\ref{fig:kapral}.
Finally, results consistent with the above model have been obtained
not only for instantaneous collision models but also for 
a system with finite range of the interaction, more precisely for 
a one-dimensional gas of particles with nearest-neighbor Coulomb interaction,
modeling a screened Coulomb interaction between electrons~\cite{Chen2015}. 
This latter model takes advantage of the recently reported 
Fourier-like behavior of thermal 
conductivity~\cite{Zhong2012,Dhar2013,Wang2013,Savin2014,Chen2014},
namely, of the appearance of a very broad range of system size where the
thermal conductivity behaves normally according to the Fourier law, i.e.,
$\kappa$ is size-independent, see Fig.~\ref{fig:fourierlike}. As a consequence, $ZT$ exhibits a rapid,
liner growth with the system size.
While the Fourier-like regime might be an intermediate (in the system size)
regime, followed by an asymptotic regime of anomalous thermal conductivity
$\kappa\sim \Lambda^{1/3}$~\cite{Lepri2003,Dhar2008}, the range of validity of such
regime may expand rapidly as an integrable limit is
approached~\cite{Chen2014}. 
We point out that it is a priori not excluded that 
there exist models
where the long-time limit $t\to\infty$ and the 
thermodynamical limit $\Lambda\to\infty$ do not commute when computing
the Drude weights.
However, numerical evidence shows that 
for the models so far considered these two limits commute
\cite{Benenti2013,Benenti2014,Chen2015}.

It is interesting to note the contrasting behavior obtained when
more than one  conserved quantities exist. For $M>1$,  in general 
${D}_{\rho\rho}{D}_{uu}  - {D}_{u\rho}^2  \ne 0$.
As a  consequence, $\det{\mathbb{L}}\sim \Lambda^2$, 
and therefore the  heat  conductivity becomes
ballistic  and $ZT$  asymptotically independent of  the  system size.  
This
situation is commonly found in  integrable systems, for which infinite
constants  of  motion  exist at the thermodynamic limit. 
For instance, in  noninteracting  systems,
momentum  conservation  implies that  all  transport coefficients  are
ballistic, thus  leading to  a constant $ZT$.  The enhancement  in the
efficiency due to the existence  of conserved quantities is limited to
systems of interacting particles.


\subsection{Breaking time-reversibility}
\label{sec:magnetic}

When time-reversal symmetry is broken, typically by a magnetic
field ${\bm B}$, Onsager-Casimir reciprocity relations no longer imply
$L_{ji}=L_{ij}$, but rather
$L_{ji}({\bm B})=L_{ij}(-{\bm B})$. While these relations
imply $\sigma({\bm B})=\sigma(-{\bm B})$ and $\kappa({\bm B})=
\kappa(-{\bm B})$, the thermopower is not bounded to be a symmetric
function under the exchange ${\bm B}\to -{\bm B}$.
This simple remark has deep consequences on thermoelectric efficiency.

\begin{figure}[!h]
\centerline{\includegraphics*[width=0.55\textwidth]{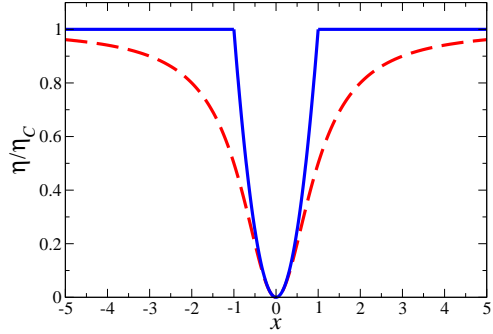}}
\caption{Efficiency $\eta$ in units of the Carnot efficiency $\eta_C$ as a function of the asymmetry parameter $x$,
with $\eta=\bar{\eta}(P_{\rm max})$ (dashed curve) and
$\eta=\bar{\eta}_{\rm max}$
(full curve). For $x=1$,
$\bar{\eta}(P_{\rm max})=\eta_C/2$ and $\bar{\eta}_{\rm max}=\eta_C$
are obtained for $y(x=1)=ZT=\infty$.}
\label{fig:magnetic}
\end{figure}

The maximum efficiency and the
efficiency at maximum power are now determined
by two parameters \cite{Benenti2011}: the asymmetry parameter
\begin{equation}
x=\frac{S({\bm B})}{S(-{\bm B})}
=\frac{S({\bm B})}{\Pi({\bm B})}\,T
\label{def:x}
\end{equation}
and the ``figure of merit''
\begin{equation}
y=
\frac{\sigma({\bm B}) S({\bm B})S(-{\bm B})}{\kappa({\bm B})}\,T
=\frac{\sigma({\bm B}) S({\bm B})\Pi({\bm B})}{\kappa({\bm B})}\ .
\end{equation}
In terms of these variables, the maximum efficiency reads
\begin{equation}
\eta_{\rm max}= \eta_C\,x\,
\frac{\sqrt{y+1}-1}{\sqrt{y+1}+1}\ ,
\label{eq:ZTx}
\end{equation}
while the efficiency at maximum power is
\begin{equation}
\eta(P_{\rm max})=
\frac{\eta_C}{2}\,\frac{xy}{2+y}\ .
\label{etawmax}
\end{equation}
In the particular case $x=1$, $y$ reduces to the $ZT$
figure of merit of the
time-symmetric case, Eq.~(\ref{eq:ZTx}) reduces to
Eq.~(\ref{eq:efficiency}),
and Eq.~(\ref{etawmax}) to Eq.~(\ref{eq:effmaxpower}).
While thermodynamics does not impose any restriction on the
attainable values of the asymmetry parameter $x$, the positivity
of entropy production implies $h(x)\le y \le 0$ if $x\le 0$ and
$0\le y \le h(x)$ if $x\ge 0$, where the function $h(x)= 4x/(x-1)^2$.
Note that $\lim_{x\to 1} h(x)=\infty$ and therefore there is no
upper bound on $y(x=1)=ZT$. For a given value of the asymmetry $x$,
the maximum (over $y$)
$\bar{\eta}(P_{\rm max})$ of $\eta(P_{\rm max})$ and the maximum
$\bar{\eta}_{\rm max}$ of $\eta_{\rm max}$ are obtained for $y=h(x)$:
\begin{equation}
\bar{\eta}(P_{\rm max})=\eta_C\,\frac{x^2}{x^2+1}\ ,
\label{eq:boundetapmax}
\end{equation}
\begin{equation}
\bar{\eta}_{\rm max}=
\left\{
\begin{array}{ll}
\eta_C\,x^2 & {\rm if}\,\, |x| \le 1\ ,
\\
\\
\eta_C & {\rm if}\,\, |x| \ge 1\ .
\end{array}
\right.
\label{eq:boundetamax}
\end{equation}
The functions $\bar{\eta}(P_{\rm max})(x)$ and
$\bar{\eta}_{\rm max}(x)$
are drawn
in Fig.~\ref{fig:magnetic}.
In the case $|x|>1$, it is in principle possible to overcome
the Curzon-Ahlborn limit within linear response
(that is, to have $\eta(P_{\rm max})>\eta_C/2$) and to reach the
Carnot efficiency, for increasingly smaller and smaller figure of merit $y$ as
the asymmetry parameter $x$ increases. The Carnot efficiency is
obtained for $\det \mathbb{L}=(L_{\rho u}-L_{u\rho})^2/4>0$ when $|x|>1$,
that is, the tight coupling condition is not fulfilled.

The output power at maximum efficiency reads
\begin{equation}
P(\bar{\eta}_{\rm max})=
\frac{\bar{\eta}_{\rm max}}{4}\frac{|L_{\rho u}^2-L_{u\rho}^2|}{L_{\rho\rho}}\,
\frac{T_\HH-T_\CC}{T^2}\ .
\end{equation}
Therefore, always within linear response,
it is allowed from thermodynamics
to have Carnot efficiency and nonzero power
simultaneously when $|x|>1$.
Such a possibility can be understood on the basis of the
following argument \cite{Saito2013,Brandner2013}. We first split 
the particle and energy currents into a
reversible part 
(which changes sign by reversing ${\bm B}\to-{\bm B}$)
and an irreversible part
(invariant with respect to the inversion ${\bm B}\to-{\bm B}$),
defined by
\begin{equation}
{\bf j}^{\rm rev}({\bm B})=\frac{\mathbb{L}({\bm B})-\mathbb{L}^t({\bm B})}{2}\,{\bf F}, 
\quad 
{\bf j}^{\rm irr}({\bm B})=\frac{\mathbb{L}({\bm B})+\mathbb{L}^t({\bm B})}{2}\,{\bf F}\ . 
\label{eq:Jrevirr}
\end{equation}
Only the irreversible part
of the currents contributes to the entropy production:
$\dot{s}={\bf j}^{\rm irr}\cdot {\bf F}=j_\rho^{\rm irr} \nabla(-\mu/T)+
j_u^{\rm irr} \nabla(1/T)$.
The reversible currents vanish for
${\bm B}=0$. On the other hand, for broken time-reversal symmetry
the reversible currents can in principle become arbitrarily large,
giving rise to the possibility of dissipationless transport.

It is interesting to compare the performances of a system as 
a thermal machine or as a refrigerator. 
For a refrigerator, the most important benchmark is
the \emph{coefficient of performance}
$\eta^{(r)}=j_q/P$ ($j_q<0$, $P<0$), given by the ratio of the heat
current extracted from the cold system over the absorbed power.
The efficiency of an ideal, dissipationless refrigerator is given by
$\eta_C^{(r)}=T_\CC/(T_\HH-T_\CC)$.
While in the time-reversal case the linear response
normalized maximum efficiency
$\eta_{\rm max}/\eta_C$
and coefficient of performance
$\eta_{\rm max}^{(r)}/\eta_C^{(r)}$
for power generation and refrigeration
coincide, this is no longer the case with broken
time-reversal symmetry.
For refrigeration
the maximum value of the coefficient of performance reads
\begin{equation}
\eta_{\rm max}^{(r)}=\eta_C^{(r)}
\,\frac{1}{x}\,\frac{\sqrt{y+1}-1}{\sqrt{y+1}+1}.
\label{eq:etarefrigeration}
\end{equation}
For small fields, $x$ is in general
a linear function of the magnetic field,
while $y$ is by construction an even function of the field.
As a consequence, a small external magnetic field either improves
power generation and
worsens refrigeration or vice-versa, while the average
efficiency
\begin{equation}
\frac{1}{2}\left[\frac{\eta_{\rm max}({\bm B})}{\eta_C}+
\frac{\eta_{\rm max}^{(r)}({\bm B})}{\eta_C^{(r)}}\right]
=\frac{\eta_{\rm max}({\bm 0})}{\eta_C}=
\frac{\eta_{\rm max}^{(r)}({\bm 0})}{\eta_C^{(r)}},
\end{equation}
up to second order corrections.
Due to the Onsager-Casimir relations, $x(-{\bm B})=1/x({\bm B})$
and therefore by inverting the direction of the magnetic field
one can improve either power generation or refrigeration.

With regard to the practical relevance of the
results presented in this section, we should note that,
as a consequence
of the symmetry properties of the scattering matrix \cite{Datta1995}
(see Sect.~\ref{sec:probe}),
in the non-interacting case
the thermopower is a symmetric function of the magnetic field,
thus implying $x=1$.
On the other hand, as we shall discuss in Sect.~\ref{sec:probe},
this symmetry may be violated when electron-phonon or electron-electron
interactions are taken into account.
Non-symmetric thermopowers have been reported
in measurements for certain orientations of
a bismuth crystal \cite{wolfe1963} and
in Andreev interferometer experiments \cite{chandrasekhar}
(for a theoretical analysis of these latter experiments see
\cite{jacquod}).

\subsection{Inelastic scattering and probe terminals}
\label{sec:probe}

\begin{figure}[!h]
\centerline{\includegraphics*[width=0.85\textwidth]{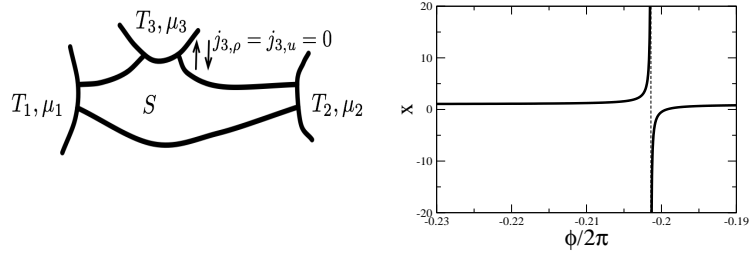}}
\caption{Left panel: schematic drawing of thermoelectric
transport, with a third terminal acting as a probe reservoir
mimicking inelastic scattering. The temperature $T_3$ and the chemical
potential $\mu_3$ of the third reservoir are such that the net
average electric and energy currents through this reservoir vanish:
$j_{3,\rho}=J_{3,u}=0$. This setup can be generalized to any number
of probe reservoirs, $k=3,...,n$, by setting
$j_{k,\rho}=J_{k,u}=0$ for all probes.
Right panel: asymmetry parameter $x$ for a three-terminal Aharonov-Bohm
interferometer, with one of the terminals acting as a probe, see 
\cite{Saito2011} for details.}
\label{fig:3ter}
\end{figure}

Inelastic scattering events like electron-phonon 
interactions, can be conveniently modeled by means
of a third terminal (or conceptual probe),
whose parameters (temperature and chemical potential)
are chosen self-consistently so that there is no net \emph{average} flux of
particles and heat between this terminal and the system
(see Fig.~\ref{fig:3ter}, left panel).
In mesoscopic physics, probe reservoirs
are commonly used to simulate phase-breaking processes in
partially coherent quantum transport, since they
introduce phase-relaxation without energy damping \cite{buttiker1988}.
The advantage of such approach lies in its simplicity and independence
from microscopic details of inelastic processes.
Probe terminals have been widely used in the literature
and proved to be useful to unveil nontrivial aspects of
phase-breaking
processes \cite{Datta1995}, heat transport and rectification
\cite{visscher,lebowitz,dhar2007,Dhar2008,lebowitz2009,pereira,segal2011,saito06},
and thermoelectric transport
\cite{Jacquet2009,Entin-Wohlman2010,Entin-Wohlman2012,Jiang2012,Jiang2013,Sanchez2011,Sanchez2011(b),Sothmann2012,Sothmann2012(b),Saito2011,Horvat2012,Segal2013,Balachandran2013,Saito2013,Bosisio2014}.

The approach can be generalized to any number $n_p$ of probe reservoirs.
We call ${\bf j}_k\equiv (j_{k,\rho},j_{k,u})^t$ the particle and energy currents
from the $k$th terminal (at temperature $T_k$ and chemical
potential $\mu_k$), with $k=3,...,n$ denoting the
$n_p=n-2$ probes.
Due to the steady-state constraints of charge and energy conservation,
$\sum_k j_{k,\rho}=\sum_k j_{k,u}=0$,
we can express, for instance, the currents from the
second reservoir as a function of the remaining $2(n-1)$ currents.
The corresponding generalized forces are given by
${\bf X}_k\equiv 
(\Delta (\mu_k/T),\Delta T_k/T^2)^t$,
with $\Delta \mu_k=\mu_k-\mu$,
$\Delta T_k= T_k-T$, $\mu=\mu_2$, and $T=T_2$.
The linear response relations between currents and thermodynamic
forces read as follows:
\begin{equation}
{\bf j}_i = \sum_{\substack{j=1
\\(j\ne 2)}}^n \mathbb{L}_{ij} {\bf X}_j,
\end{equation}
where $\mathbb{L}_{ij}$ are $2\times 2$ matrices, so that 
the overall Onsager matrix $\mathbb{L}$ has size $2(n-1)$.
We then impose the condition of zero average currents through
the probes, $j_{k,\rho}=j_{k,u}=0$ for $k=3,...,n$ to reduce
the Onsager matrix to a $2\times 2$ matrix $\mathbb{L}'$
connecting the fluxes ${\bf j}_1$ through the first reservoir
and the conjugated forces ${\bf X}_1$ as ${\bf j}_1=\mathbb{L}'{\bf X}_1$.
The reduced matrix $\mathbb{L}'$ fulfills the Onsager-Casimir
relations and represents the Onsager matrix for two-terminal
inelastic transport modeled by means of probe reservoirs.
The transport coefficients and the thermodynamic efficiencies can then 
be computed in the usual way from the reduced matrix $\mathbb{L}'$.

The particle and
energy currents can be conveniently computed, for any number of probes, by
means of the multi-terminal Landauer-B\"uttiker formula \cite{Datta1995}:
\begin{equation}
j_{k,\rho}=\frac{1}{h}\int_{-\infty}^{\infty}
d\epsilon 
\sum_l[\tau_{l\leftarrow k}(\epsilon)f_k(\epsilon)
-\tau_{k\leftarrow l}(\epsilon)f_l(\epsilon)],
\end{equation}
\begin{equation}
j_{k,u}=\frac{1}{h}\int_{-\infty}^{\infty}
d\epsilon \epsilon 
\sum_l[\tau_{l\leftarrow k}(\epsilon)f_k(\epsilon)
- \tau_{k\leftarrow l}(\epsilon)f_l(\epsilon)],
\end{equation}
where $\tau_{l\leftarrow k}(\epsilon)$ is the transmission probability
from terminal $k$ to terminal $l$ at the energy $\epsilon$.
Charge conservation and the requirement
of zero current at zero bias impose
\begin{equation}
\sum_k\tau_{k\leftarrow l}=
\sum_{k}\tau_{l\leftarrow k}=M_l,
\label{eq:multiterminal}
\end{equation}
with $M_l$ being the number of modes in the lead $l$. Moreover, in
the presence of a magnetic
field ${\bm B}$ we have
\begin{equation}
\tau_{k\leftarrow l}({\bm B})=\tau_{l\leftarrow k}(-{\bm B}).
\label{eq:taumag}
\end{equation}
The last relation is a consequence of the unitarity of the
scattering matrix $\mathbb{S}({\bm B})$ that relates the
outgoing wave amplitudes to the incoming wave amplitudes at the
different leads. The time-reversal invariance of unitary
dynamics leads to $\mathbb{S}({\bm B})=\mathbb{S}^t(-{\bm B})$,
which in turn implies (\ref{eq:taumag}) \cite{Datta1995}.
In the two-terminal case, Eq.~(\ref{eq:multiterminal}) means
$\tau_{1\leftarrow 2}=\tau_{2\leftarrow 1}$. Hence, we can
conclude from this relation and Eq.~(\ref{eq:taumag}) that
$\tau_{2\leftarrow 1}({\bm B})=\tau_{2\leftarrow 1}(-{\bm B})$,
thus implying that the Seebeck coefficient is a symmetric
function of the magnetic field.

Probe terminals
can break the symmetry of the Seebeck coefficient.
We can have $S(-{\bm B})\ne S({\bm B})$, that is, 
$L'_{12}\ne L'_{21}$ in the reduced Onsager matrix
$\mathbb{L}'$. Arbitrarily large values of the asymmetry
parameter $x=S({\bm B})/S(-{\bm B})$
were obtained in \cite{Saito2011} (see Fig.~\ref{fig:3ter}, right panel) by means
of a three-dot Aharonov-Bohm interferometer model.
The asymmetry was found also for 
chaotic cavities,
ballistic microjunctions \cite{Sanchez2011(b)}, and
random Hamiltonians drawn from the Gaussian unitary
ensemble \cite{Balachandran2013}, and also in the framework
of classical physics, for a three-terminal deterministic
railway switch transport model \cite{Horvat2012}.
In the latter model, only the values
zero and one are allowed for the transmission functions
$\tau_{j\leftarrow i}(\epsilon)$, i.e., $\tau_{j\leftarrow i}(\epsilon)=1$
if particles injected from terminal $i$ with energy $\epsilon$
go to terminal $j$ and $\tau_{j\leftarrow i}(\epsilon)=0$ is such
particles go to a terminal other than $j$. The transmissions
$\tau_{j\leftarrow i}(\epsilon)$ are piecewise constant
in the intervals $[\epsilon_i,\epsilon_{i+1}]$, $(i=1,2,...)$, with switching
$\tau_{j\leftarrow i}=1\to 0$ or viceversa possible at the
threshold energies $\epsilon_i$, with the constraints (\ref{eq:multiterminal})
always fulfilled.

\begin{figure}[!h]
\centerline{\includegraphics*[width=0.9\textwidth]{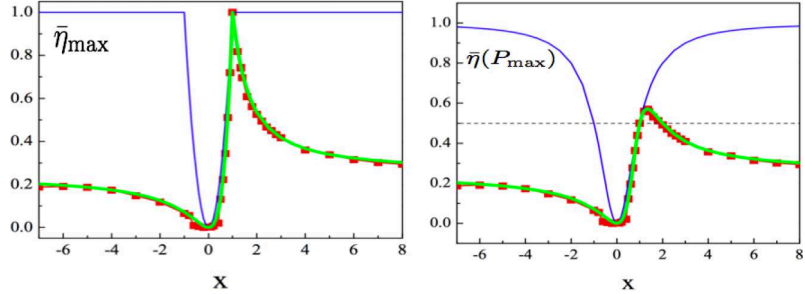}}
\caption{Maximum efficiency $\bar{\eta}_{\rm max}$  (left panel) and
efficiency at maximum power
$\bar{\eta}(P_{\rm max})$ (right panel), both in units of $\eta_C$.
Upper curves correspond to the thermodynamics bounds \cite{Benenti2011},
lower curves to the more restrictive bounds \cite{Saito2013} from 
the unitarity of the scattering matrix for three terminals,
squares are obtained from a transmission model whose details are described
in \cite{Balachandran2013}. Dotted-dashed line corresponds to the Curzon-Ahlborn 
linear-response limit $\eta_{C}/2$.  Note that such limit is exceeded 
in the interval [1,2] with the transmission model.}
\label{fig:3terbounds}
\end{figure}

In all the above instances, it was not possible to find at the same
time large values of asymmetry parameter (\ref{def:x}) and high thermoelectric efficiency. 
Such failure was explained by \cite{Saito2013} and is generic
for non-interacting three-terminal systems. In that case, when
the magnetic field ${\bm B}\ne 0$, current conservation, which is
mathematically expressed by unitarity of the scattering matrix
$\mathbb{S}$, imposes bounds on the Onsager matrix stronger than
those derived from the positivity of entropy production.
As a consequence, Carnot efficiency can be achieved
in the three-terminal setup only in the symmetric case
$x=1$. On the other hand, the Curzon-Ahlborn linear response
bound, $\eta_C/2$, for the efficiency at maximum power
can be overcome for moderate asymmetries, $1<x<2$, with a
maximum of $4\eta_C/7$ at $x=4/3$.
The bounds obtained by
\cite{Saito2013} are in practice saturated in a
quantum transmission model reminiscent of the
above described railway switch model \cite{Balachandran2013}
(see Fig.~\ref{fig:3terbounds}).
The generic multi-terminal case
was also discussed for
noninteracting electronic transport \cite{Brandner2013}.
By increasing the number $n_p$ of probe terminals,
the constraint from current conservation on the maximum efficiency
and the efficiency at maximum power becomes weaker. However, the bounds
(\ref{eq:boundetapmax}) and (\ref{eq:boundetamax})
from the second law of
thermodynamics are saturated only in the limit $n_p\to\infty$.
Moreover, numerical evidence suggests that the power 
vanishes when the maximum efficiency is approached~\cite{Brandner2015}.
It is an interesting open question whether
similar bounds on efficiency, tighter
that those imposed by the positivity of entropy production,
exist in more general transport models for interacting systems.

Finally, we point out that in a genuine multi-terminal device
all terminals should be treated on equal footing, without 
necessarily declaring some of them as probes. 
First investigations for a generic three-terminal system
have shown that in some instances the coupling to a third terminal can
improve both the extracted power and the efficiency
of a thermoelectric device~\cite{Mazza2014}.
Moreover, with three terminals one can separate the currents,
with charge and heat flowing to different reservoirs.
As a result, it is possible to violate in a controlled fashion the
Wiedemann-Franz law, greatly enhancing thermoelectric performances \cite{Mazza2015}.

\section{Concluding remarks}

In this chapter we have discussed several microscopic mechanisms for the 
design of a thermal rectifier and the increase of the efficiency of thermoelectric 
energy conversion.  
Although not intuitive, solid-state thermal rectifiers do exist and
there have already been the first experimental implementations. 
With regard to thermoelectricity, basic concepts to improve the 
efficiency have been identified: energy filtering for non-interacting systems
and momentum conservation in non-integrable interacting systems.

Several questions remain open. An important point for thermal rectification is
the need to have a strongly temperature dependent thermal conductivity. 
Some ideas have already been explored, but the microscopic theory is still incomplete.
It appears promising in this connection to work in the vicinity of a 
structural phase transition.
Moreover, the above discussed rectifiers are based on insulating materials. 
It would be interesting, in order to combine thermal rectification with thermoelectric
power generation or cooling, to include and understand the role of 
mobile charge carriers.
Recent experimental investigations are moving forward in this direction
\cite{Giazotto2014}.

In spite of the long history of thermoelectricity, from the viewpoint of statistical 
physics the theory of the coupled transport of heat and charge is still in 
its infancy. With regard to the challenging problem of improving the 
efficiency of heat to work conversion, for non-interacting systems we 
have a quite complete theoretical picture and understand the 
limitations imposed by nature (notably, the Wiedemann-Franz law).
On the other hand, the understanding of general mechanisms connected 
to strongly interacting systems, for which the Wiedemann-Franz law does
not apply, are only beginning to emerge. In particular, regimes near 
electronic phase transitions might be favorable for 
thermoelectric conversion \cite{Vining1997,Ouerdane2014}. 
A deeper understanding of the 
nonlinear regime is also needed \cite{jacquod13,sanchez13,whitney13}, 
since, as observed experimentally in mesoscopic devices \cite{matthews13},
the Onsager-Casimir reciprocity relations break down and this 
fact could in principle allow for improved thermoelectric efficiencies. 
Furthermore, in the nonlinear regime rectification effects occur and 
their impact on thermoelectricity is still not well understood. 

\begin{acknowledgement}
G.B. and G.C. acknowledge the support by MIUR-PRIN.
C.M.-M. acknowledges the support by the Spanish MICINN
  grant MTM2012-39101-C02-01.
\end{acknowledgement}
%
%
%
%


\end{document}